\documentclass[12pt,a4paper]{article}
\usepackage{amsfonts}
\usepackage{amsmath}
\usepackage{amssymb}
\usepackage{latexsym}

\usepackage[dvips]{graphicx}

\numberwithin{equation}{section}

\newcommand{\lanln}[1]{$\langle$\texttt{arXiv:#1}$\rangle$}

\newcommand{\BbbR}{\mathbb{R}}

\newcommand{\auxeps}{\eta} 

\newcommand{\vsigma}{\mathsf{q}} 

\newcommand{\scri}{\mathcal{I}}

\title{How often does the Unruh-DeWitt detector click? 
\\
Regularisation by a spatial profile.}

\author{Jorma Louko\thanks{jorma.louko@nottingham.ac.uk}
\ and
Alejandro Satz\thanks{pmxas3@nottingham.ac.uk}
\\
\noalign{\vspace{3ex}}
\small{\it School of Mathematical Sciences,
University of Nottingham,}\\
\small{\it Nottingham NG7 2RD, UK}
\\
\noalign{\vspace{1ex}}\\
\small{(Revised September 2006)}
\\
\noalign{\vspace{1ex}}
\small{\lanln{gr-qc/0606067}}
\\
\noalign{\vspace{3ex}}
\small{Class.\ Quantum Grav.\ \textbf{23} (2006) 6321--6343}
}

\date{}

\begin{document}

\maketitle

\begin{abstract}

We analyse within first-order perturbation theory the instantaneous
transition rate of an accelerated Unruh-DeWitt particle detector whose
coupling to a massless scalar field on four-dimensional Minkowski
space is regularised by a spatial profile. For the Lorentzian profile
introduced by Schlicht, the zero-size limit is computed explicitly and
expressed as a manifestly finite integral formula that no longer
involves regulators or limits. The same transition rate is obtained
for an arbitrary profile of compact support under a modified
definition of spatial smearing. Consequences for the asymptotic
behaviour of the transition rate are discussed. A~number of stationary
and nonstationary trajectories are analysed, recovering in particular
the Planckian spectrum for uniform acceleration.

\end{abstract}

\newpage

\section{Introduction}

The fact that a uniformly accelerated observer 
in Minkowski space perceives the Minkowski vacuum as a 
thermal state with a temperature proportional to the 
acceleration is called the Unruh effect in honour of its 
discoverer \cite{unruh}. It is the simplest among 
the phenomena linking thermal effects to spacetime horizons. 
Other examples 
of such phenomena are thermalisation 
of de~Sitter space as seen by inertial observers 
\cite{gibb-haw:dS} 
and the celebrated discovery by Hawking of the thermal 
radiation surrounding black holes~\cite{hawking}.

A conceptually straightforward way to address the Unruh 
effect is in terms of a particle detector model, consisting of 
a quantum system with 
discrete energy levels and a weak coupling 
to the quantum 
field. In generic motion the detector will undergo transitions, which 
can be interpreted 
as due to absorption or emission of field quanta; the particle 
content of the 
field is thus defined operationally by reference to the 
measurable excitations or de-excitations of 
the detector. The simplest model, 
originated by Unruh \cite{unruh} and DeWitt~\cite{deWitt}, 
involves a linear coupling of the detector's 
monopole moment to 
the field at the detector's position. 
For the uniformly accelerated detector in 
Minkowski vacuum this model
exhibits the Planckian spectrum for the 
(infinite) total transition probability 
divided by the (infinite) total proper time
\cite{unruh,deWitt,byd}, 
under certain technical assumptions on handling the infinities. 

If the detector is allowed to move on an arbitrary (timelike)
trajectory, the notion of transition probability becomes more
subtle. As we will review in section~\ref{sec:detectormodels}, the
formal first-order perturbation theory expression for the
instantaneous transition rate involves the distribution-valued
Wightman function of the quantum field in a way whose interpretation
is ambiguous. Schlicht \cite{schlicht} observed that when the Wightman
function is represented by the `standard' $i\epsilon$ regularisation
in a given Lorentz frame, the resulting instantaneous transition rate
for a uniformly accelerated detector that has been switched on in the
asymptotic past depends on the proper time of the detector; a result that
breaks Lorentz invariance and 
appears thus physically incorrect. 
Schlicht also showed that when the
detector response is regularised by first giving the detector a
specific spatial profile and then taking the point-like limit, the
result is equivalent to giving the Wightman function along the
detector world line a non-standard regularisation. This nonstandard
regularisation yields physically reasonable results for a number of
trajectories, including the time-independent Planckian transition rate
for the uniformly accelerated detector
\cite{schlicht,Schlicht:thesis}. P.~Langlois \cite{Langlois} observed
that this regularisation can be alternatively interpreted as an
exponential frequency cut-off in the detector's instantaneous rest
frame, rather than in a fixed Lorentz frame.

While Schlicht's regularisation of the detector by a spatial profile
thus appears physically reasonable, the results in
\cite{schlicht,Schlicht:thesis} for non-inertial motion rely on a
specific choice for the profile function: the Lorentzian function, given
below in our~(\ref{lorentzian}). (For inertial motion, it was shown in
\cite{Schlicht:thesis} that the response in Minkowski vacuum is the
same for any spherically symmetric profile.) Further, the final
transition rate formula in \cite{schlicht,Schlicht:thesis} involves an
integral of the \emph{regularised\/} correlation function, and the
regulator may be taken to zero only after the integration. Such a
formula can be readily applied to specific trajectories, at least
numerically, but it does not provide a transparent starting point for
extracting analytically properties of interest, such as asymptotic
behaviour at large/small frequency or at early/late proper times.

The purpose of this paper is to address the above two questions. We
first compute the zero-size limit of Schlicht's transition rate
explicitly and show that it can be written as a manifestly finite
integral formula that no longer involves regulators or limits. The
result holds both for a detector switched on at a finite time and for
a detector switched on in the asymptotic past, in the latter case
subject to certain asymptotic conditions on the trajectory. We also
show that the transition rate obtained from the `standard' $i\epsilon$
regularisation of the Wightman function in a given Lorentz frame
differs by an additive, Lorentz-noninvariant term that is independent
of the frequency and local as a function along the trajectory. This
Lorentz-noninvariant term agrees with the analytic and numerical
observations made in the special case of uniform acceleration
in~\cite{schlicht}.

Second, we discuss a spatially extended detector model that
uses a modified definition of spatial smearing. 
Subject to mild technical conditions on both the 
detector profile and the trajectory, we show that 
the transition rate in this model is 
\emph{independent\/} of the spatial profile
and coincides with that obtained from 
the (unmodified) smearing with the Lorentzian profile function. 

Third, we use our transition rate formula to make certain observations
on the parity and falloff properties of the detector
response. Finally, we examine a number of specific trajectories,
including all the stationary worldlines in Minkowski space and a
selection of nonstationary worldlines with interesting asymptotics. We
recover in particular the Planckian spectrum for uniform acceleration,
and for a motion interpolating between inertial and uniformly
accelerated motion we obtain an appropriately interpolating transition
rate.

We begin in section \ref{sec:detectormodels} with a brief review of
the Unruh-DeWitt particle detector and its regularisation by a spatial
profile. The zero-size limit of the transition rate regularised by the
Lorentzian profile function is computed in
section~\ref{sec:lor-limit}. The modified spatially extended detector
model is analysed in section~\ref{sec:generalprofile}. Section
\ref{sec:applications} discusses the parity and falloff properties of
the transition rate and presents the applications to specific
trajectories. The results are summarised and discussed in
section~\ref{sec:conclusions}. The proofs of certain technical results
are deferred to two appendices.

We work in four-dimensional 
Minkowski spacetime with metric signature $({-} {+} {+} {+})$, 
and in units in which $\hbar= c=1$. 
Boldface letters denote spatial 
three-vectors and sans-serif letters spacetime four-vectors.
The Euclidean scalar product of three-vectors 
$\mathbf{k}$ and $\mathbf{x}$ 
is denoted by 
$\mathbf{k}\cdot \mathbf{x}$, 
and the Minkowski scalar product of four-vectors 
$\mathsf{k}$ and $\mathsf{x}$
is denoted by 
$\mathsf{k}\cdot \mathsf{x}$. 
$O(x)$ denotes a quantity for which 
$O(x)/x$ is bounded as $x\to0$ 
and $O(1)$ 
a quantity that remains bounded when the 
parameter under consideration approaches zero.

\section{Particle detector models}
\label{sec:detectormodels}

We begin by considering a detector that consists of an 
idealised, pointlike atom with two energy levels, 
denoted by $\vert0\rangle_d$ and $\vert1\rangle_d$, 
which are eigenstates of the atomic Hamiltonian $H_d$ 
with respective eigenvalues $0$ and $\omega$, $\omega\ne0$. 
The detector is coupled to the 
real, massless scalar field $\phi$ 
at the position of the detector, 
with an interaction Hamiltonian of the form
$H_{\mathrm{int}}=c \chi(\tau) \mu(\tau)\phi 
\bigl(\mathsf{x}(\tau) \bigr)$, 
where $c$ is a coupling constant, 
$\mu(\tau)$ is the atom's monopole moment operator and 
$\mathsf{x}(\tau)$ is the spacetime position of the 
atom parametrised by its proper time~$\tau$. 
$\chi(\tau)$ is a smooth switching function, 
positive during the interaction and vanishing 
before and after the interaction. 

Suppose that before the interaction the detector is in the state 
$\vert0\rangle_d$ and the field in the state $\vert A\rangle$. 
After the interaction has taken place, 
the detector may have a nonzero probability 
to be in the state 
$\vert1\rangle_d$: for $\omega>0$ (respectively $\omega<0$), 
the transition of the detector 
is interpreted as the absorption (emission) 
of a particle of energy~$\omega$. 
In first-order perturbation theory 
in the coupling constant~$c$, this 
probability is \cite{unruh,deWitt,byd,wald-smallbook,Junker}
\begin{equation}
\label{eq:total-probability}
P(\omega)
=
c^2
\, 
{\bigl\vert
{}_d\langle0\vert\mu(0)\vert1\rangle_d\bigr\vert}^2
F(\omega)
\ , 
\end{equation}
where the response function 
$F(\omega)$ is given by 
\begin{equation}
\label{defresponse}
F(\omega)= \int_{-\infty}^{\infty}
\mathrm{d}\tau' \int_{-\infty}^{\infty}\mathrm{d}\tau'' 
\, 
\mathrm{e}^{-i\omega(\tau'-\tau'')}
\, 
\chi(\tau')\chi(\tau'')
\, 
\langle A\vert 
\phi\bigl(\mathsf{x}(\tau')\bigr)
\phi\bigl(\mathsf{x}(\tau'')\bigr)
\vert A\rangle
\ . 
\end{equation}
The response function $F(\omega)$ 
encodes the part of the probability that depends on
the trajectory but not on the detector's internal properties, 
while the prefactor that involves the matrix element 
${}_d\langle0\vert\mu(0)\vert1\rangle_d$ 
depends on the detector's 
internal properties but not on the trajectory. 

We now specialise to four-dimensional Minkowski spacetime $M^4$ 
and take the quantum state $\vert A\rangle$ 
of the field to be the Minkowski vacuum, denoted by $\vert 0\rangle$. 
The Wightman function 
$\langle 0\vert \phi(\mathsf{x})\phi(\mathsf{x}')\vert 0\rangle$ 
is then a well-defined distribution on $M^4 \times M^4$, 
and its pull-back to the detector world line, the correlation function 
\begin{equation}
\label{eq:W} 
W(\tau',\tau'') := 
\langle 0\vert 
\phi \bigl(\mathsf{x}(\tau') \bigr)
\phi \bigl(\mathsf{x}(\tau'') \bigr)
\vert 0\rangle
\ , 
\end{equation}
is a well-defined distribution on 
$\BbbR \times \BbbR$ \cite{Junker}. 
As long as the switching function $\chi(\tau)$ is assumed 
smooth and of compact support, formula 
(\ref{defresponse}) gives therefore an unambiguous definition to the 
response function~$F(\omega)$, and the probability 
(\ref{eq:total-probability}) 
is that of observing the detector in the state 
$\vert1\rangle_d$ 
after the interaction has ceased. 

We wish to address a related but subtly different question
\cite{schlicht,finitetime1,finitetime2,finitetime3}: 
What is the probability of finding the detector 
in the state $\vert1\rangle_d$ 
\emph{while the interaction is still switched on?\/} 
Proceeding for the moment formally, 
the answer should be obtained from 
(\ref{eq:total-probability}) 
and 
(\ref{defresponse}) 
by introducing in the switching function a sharp cut-off
at the proper time $\tau$ at which the detector is observed, 
$\chi(\tau') \to \chi(\tau') \Theta(\tau - \tau')$, 
where $\Theta$ is the Heaviside function. 
The modified response function 
$F_\tau(\omega)$ then reads 
\begin{equation}
\label{defresponse-tau}
F_\tau(\omega)= 
\int_{-\infty}^{\tau}
\mathrm{d}\tau' \int_{-\infty}^{\tau}\mathrm{d}\tau'' 
\, 
\mathrm{e}^{-i\omega(\tau'-\tau'')}
\, 
\chi(\tau')\chi(\tau'')
\, 
W(\tau',\tau'')
\ . 
\end{equation}
The physically interesting quantity in this situation is
the instantaneous transition rate 
$\dot{F}_{\tau}(\omega)$, 
where the overdot denotes derivative with respect to~$\tau$. 
Up to a proportionality constant, 
$\dot{F}_{\tau}(\omega)$ 
gives the number of transitions 
to the state $\vert1\rangle_d$ 
per unit proper time in an ensemble of 
identical detectors following the trajectory~$\mathsf{x}(\tau)$. 
Differentiating (\ref{defresponse-tau}), using the identify 
$W(\tau,\tau')= \overline{W(\tau',\tau)}$ 
and performing a change of variables in the remaining integral, 
we find 
\begin{equation}
\label{defexcitation}
\dot{F}_{\tau} (\omega)
=
2 \, 
\chi(\tau) 
\, 
\mathrm{Re}
\int_{0}^{\infty}
\mathrm{d}s
\,\,
\mathrm{e}^{-i\omega s}
\, 
W(\tau,\tau-s)
\, 
\chi(\tau-s)
\ . 
\end{equation} 
If the switching function is taken to have a sharp
switch-on at the initial time $\tau_0$
and to be unity thereafter, 
$\chi(\tau') \to \Theta(\tau' - \tau_0)$, 
(\ref{defexcitation}) becomes 
\begin{equation}
\label{defexcitation-sharp}
\dot{F}_{\tau} (\omega)
=
2 \, 
\mathrm{Re}
\int_{0}^{\tau - \tau_0}
\mathrm{d}s
\,\,
\mathrm{e}^{-i\omega s}
\, 
W(\tau,\tau-s)
\ , 
\ \ \tau> \tau_0
\ . 
\end{equation} 
For a detector switched on in the distant past, 
$\tau_0 \to -\infty$, 
we obtain 
\begin{equation}
\label{defexcitation-sharp-infty}
\dot{F}_{\tau} (\omega)
=
2 \, 
\mathrm{Re}
\int_{0}^{\infty}
\mathrm{d}s
\,\,
\mathrm{e}^{-i\omega s}
\, 
W(\tau,\tau-s)
\ . 
\end{equation} 
As stressed by Schlicht \cite{schlicht} in the context of formula 
(\ref{defexcitation-sharp-infty}), 
all the expressions 
(\ref{defresponse-tau})--(\ref{defexcitation-sharp-infty}) 
are \emph{causal\/}: 
The probability and the instantaneous 
transition rate are given by integrals over the 
\emph{past\/} 
of the detector trajectory. 
There is no need to specify what the trajectory will do after the moment
the detector is read (cf.~\cite{akhmedov}). 

Now, the difficulty 
with formulas 
(\ref{defresponse-tau})--(\ref{defexcitation-sharp-infty}) 
is that each of them involves 
the correlation function $W$ in a way that is 
ambiguous. 
As $W$ is a distribution on 
$\BbbR \times \BbbR$ \cite{Junker}, 
rather than a function, the problem first arises in 
(\ref{defresponse-tau}) because 
$W$ is integrated against a function that is not smooth but has a 
discontinuity at the observation time. 
In (\ref{defexcitation-sharp}) and 
(\ref{defexcitation-sharp-infty}) the problem 
is compounded respectively by the 
discontinuity of the switching function 
at the sharp switch-on time 
and the noncompact support 
of the switching function. 

To see the ambiguity explicitly, 
recall that the formal 
mode sum expression for the 
Wightman function reads 
\begin{equation}
\label{intovermodes}
\langle 0\vert \phi(\mathsf{x})
\phi(\mathsf{x}')\vert 0\rangle
=
\frac{1}{(2\pi)^3}
\int
\frac{\mathrm{d^3}k}{2\vert\mathbf{k}\vert}
\mathrm{e}^{i\mathsf{k} \cdot (\mathsf{x}-\mathsf{x}')}
\ , 
\end{equation}
where $\mathsf{k} = (\vert\mathbf{k}\vert, \mathbf{k})$. 
The usual way to regularise this mode sum is by the frequency cut-off 
$\mathsf{x-x'} \to 
\mathsf{x-x'} -i\epsilon \partial_t$, 
where $\epsilon >0$, leading to the representation 
\cite{byd} 
\begin{equation}
\label{tradWightman}
\langle 0\vert \phi(\mathsf{x})\phi(\mathsf{x}')\vert 0\rangle
=
\lim_{\epsilon \to 0_+} 
\frac{-1}{4\pi^2}\frac{1}{{(t-t'-i\epsilon)}^2
- 
{\vert\mathbf{x}-\mathbf{x'}\vert}^2}
\ . 
\end{equation}
Assuming 
$\mathbf{x}-\mathbf{x'} \ne \mathbf{0}$, 
expanding  
(\ref{tradWightman}) into partial 
fractions and taking the limit in the sense of 
one-dimensional Hilbert transforms yields 
\begin{align}
\langle 0\vert \phi(\mathsf{x})\phi(\mathsf{x}')\vert 0\rangle
\notag 
& = 
\frac{1}{8\pi^2\vert\Delta\mathbf{x}\vert}
\left\{ 
P\left( \frac{1}{\Delta t-\vert \Delta \mathbf{x}\vert}\right)
+
P\left( \frac{1}{\Delta t+\vert \Delta \mathbf{x}\vert}\right) 
\right. 
\notag 
\\
& 
\hspace{15ex}
\left. 
\vphantom{\left( \frac{1}{\Delta t-\vert \Delta \mathbf{x}\vert}\right)}
+i\pi\bigl[ 
\delta(\Delta t+\vert \Delta \mathbf{x}\vert)-\delta(\Delta
t-\vert \Delta \mathbf{x}\vert)
\bigr] 
\right\} 
\ , 
\label{dist}
\end{align}
where 
$\Delta t := t - t'$, 
$\Delta \mathbf{x} := 
\mathbf{x}-\mathbf{x'}$ and 
$P$ stands for the Cauchy principal value. However, 
inserting the correlation function given by 
(\ref{eq:W}) with 
(\ref{dist}) 
into (\ref{defexcitation}) 
gives an integral whose interpretation is ambiguous
because $\Delta t$ and $\Delta \mathbf{x}$ both vanish at $s=0$. 

One is therefore led to seek a meaning for the 
instantaneous transition rates
in (\ref{defexcitation})--(\ref{defexcitation-sharp-infty}) 
by first regularising~$W$, then doing the integral over $s$ 
and at the end removing the regulator. 
However, 
there is now an issue as to which regularisation to use. 
The Wightman function is a well-defined, Lorentz-invariant distribution, 
and it can be defined as 
\begin{equation}
\label{covWightman}
\langle 0\vert \phi(\mathsf{x})\phi(\mathsf{x}')\vert 0\rangle
=
\lim_{\epsilon \to 0_+} 
\frac{-1}{4\pi^2}\frac{1}{{(t-t')}^2
- 
{\vert\mathbf{x}-\mathbf{x'}\vert}^2
- i \epsilon 
\bigl[ 
T(\mathsf{x}) - T(\mathsf{x}')
\bigr] 
- \epsilon^2}
\ , 
\end{equation}
where $T$ is \emph{any\/} global time function that increases to the
future~\cite{kay-wald}. The usual representation (\ref{tradWightman})
is obtained with the choice $T(\mathsf{x}) = t$ in a specific
Lorentz frame. But when the Wightman function, or the correlation
function $W$ obtained from it, is used outside the distributional
setting, there is no a priori guarantee for different time functions
in (\ref{covWightman}) to lead to the same results, nor is there a
guarantee for the results to be Lorentz invariant. The transition rate
formulas (\ref{defexcitation-sharp}) and
(\ref{defexcitation-sharp-infty}) are a case in point: Schlicht
\cite{schlicht} observed that when used
in~(\ref{defexcitation-sharp-infty}), the usual Wightman function
regularisation (\ref{tradWightman}) yields for the uniformly
accelerated detector a result that is not invariant under the
Lorentz boosts that leave the trajectory invariant. We show in
appendix \ref{app:ieps} that the only trajectories for which the usual
Wightman function regularisation (\ref{tradWightman}) yields a
Lorentz-invariant result for the transition rates
(\ref{defexcitation-sharp}) and (\ref{defexcitation-sharp-infty}) are
the inertial trajectories. If we insist on a Lorentz-invariant
transition rate, the usual regularisation (\ref{tradWightman}) will
therefore not do.

Schlicht \cite{schlicht} proposed to regularise the transition rate by
giving the detector a finite spatial extent in its instantaneous rest
frame. The idea is that the detector is not 
coupled to the 
field operator on the detector world line, 
$\phi
\bigl(\mathsf{x}(\tau) \bigr)$, 
but instead to the spatially smeared field
operator,
\begin{equation}
\label{smeared}
\phi_f(\tau) : =
\int \mathrm{d}^3 \xi
\,
f_\epsilon (\boldsymbol{\xi})
\, 
\phi 
\bigl( \mathsf{x}(\tau,\boldsymbol{\xi}) \bigr)
\ , 
\end{equation}
where $(\tau,\boldsymbol{\xi})=(\tau,\xi^1,\xi^2,\xi^3)$ is a set of
Fermi-Walker coordinates associated with the trajectory:
$\mathsf{x}(\tau,\boldsymbol{\xi}) =
\mathsf{x}(\tau)+\xi^i\mathsf{e}_i
=
\mathsf{x}(\tau)+\xi^1 \mathsf{e}_{1}
+\xi^2 \mathsf{e}_{2}+ \xi^3 \mathsf{e}_{3}$,
where $\mathsf{e}_i$ 
are three unit vectors that together with the
velocity $\dot{\mathsf{x}}$ 
form an orthonormal tetrad, 
Fermi-Walker transported with the
motion~\cite{mtw}. 
Note that the $\boldsymbol{\xi}$-coordinates 
parametrise the hyperplane
orthogonal to the velocity at each moment of proper time.
The profile function $f_\epsilon$ is assumed normalised so that 
$\int d^3\xi \, f_\epsilon(\boldsymbol{\xi}) = 1$, 
to depend on a positive parameter $\epsilon$ 
so that $\lim_{\epsilon\to0}
f_\epsilon(\boldsymbol{\xi})
=
\delta(\boldsymbol{\xi})$, and to have the scaling property 
\begin{equation}
\label{eq:profile-scaling}
f_\epsilon(\boldsymbol{\xi})
= 
\epsilon ^{-3} 
f (\boldsymbol{\xi}/\epsilon)
\ . 
\end{equation}
The function $f$ thus gives the detector's ``shape," 
which is rigid 
in the detector's instantaneous rest frame, 
and the positive parameter 
$\epsilon$ determines the detector's 
``size.'' 
If the transition rate of a detector 
smeared in this way is well defined 
and has a well-defined limit as 
$\epsilon\to0$, this limit can then be 
understood as the transition rate of a pointlike detector. 

The correlation function for the smeared field operator 
(\ref{smeared}) is defined by 
\begin{align}
\label{eq:smeared-correlator}
W_\epsilon(\tau, \tau') 
:= 
\langle0\vert\phi_f(\tau)\phi_f(\tau')\vert0\rangle
\ . 
\end{align}
When the function $f$ in (\ref{eq:profile-scaling})
is the Lorentzian function,
\begin{equation}
\label{lorentzian}
f(\boldsymbol{\xi})
=
\frac{1}{\pi^2}
\frac{1}{\bigl({\vert\boldsymbol{\xi}\vert}^2+1\bigr)^2}
\ , 
\end{equation}
Schlicht \cite{schlicht} shows that 
\begin{equation}
\label{corrSchlicht}
W_\epsilon (\tau,\tau')
=
\frac{1}{4\pi^2}\frac{1}{\bigl( \mathsf{x}-\mathsf{x}'
-i\epsilon 
(\dot{\mathsf{x}}+\dot{\mathsf{x}}') 
\bigr)^2}
\ , 
\end{equation}
where the unprimed and primed quantities on the right-hand side 
are evaluated respectively at $\tau$ and~$\tau'$. 
Note that $W_\epsilon$ (\ref{corrSchlicht}) 
is manifestly Lorentz covariant because of the four-velocities
appearing with~$\epsilon$. 
Schlicht examines \cite{schlicht,Schlicht:thesis} 
the transition rate 
(\ref{defexcitation-sharp-infty}) for a detector 
switched on in the asymptotic past using 
$W_\epsilon$ (\ref{corrSchlicht}) 
for a number of trajectories and finds results that 
appear physically sensible. In particular, 
the transition rate of a uniformly accelerated detector 
has the expected $\tau$-independent 
Planckian spectrum 
of the Unruh effect, 
\begin{equation}
\label{planck}
\dot{F}_{\tau}(\omega)=\frac{\omega}{2\pi}
\frac{1}{\mathrm{e}^{2\pi\omega/a}-1}
\ , 
\end{equation}
where $a$ is the acceleration. 
Schlicht's results have been generalised by 
P.~Langlois \cite{Langlois,Langlois-thesis} to a variety of
situations, including Minkowski 
space in an arbitrary number of dimensions, quotients of
Minkowski space under discrete isometry groups, the massive scalar
field, the massless Dirac field 
and certain curved spacetimes. 

These results of Schlicht and Langlois rely on the choice of the
Lorentzian profile function, given in four dimensions by
(\ref{lorentzian}) and in other dimensions by the appropriate
generalisation~\cite{Langlois}. For the inertial detector switched on
in the asymptotic past, it is shown in \cite{Schlicht:thesis} that all
profile functions of the form (\ref{eq:profile-scaling}) with a
spherically symmetric $f$ give the transition rate
$-\omega\Theta(-\omega)/(2\pi)$: This agrees with the results
established in~\cite{byd}. We shall address the profile-dependence of
the transition rate in section
\ref{sec:generalprofile} under a 
modified definition of the spatial smearing.

\section{Zero-size limit with the Lorentzian profile function}
\label{sec:lor-limit}

In this section we evaluate the zero-size limit of the instantaneous
transition rate for the regularised correlation
function~(\ref{corrSchlicht}), obtained in
\cite{schlicht} from spatial smearing with the Lorentzian profile
function~(\ref{lorentzian}). 
We first address a detector switched on 
at a finite proper time and then a detector 
switched on in the asymptotic past.

\subsection{Sharp switch-on}
\label{subsec:lor-sharp}

We denote by $\tau_0$ the moment of the sharp switch-on
and by $\tau$ the moment of observation, assuming $\tau> \tau_0$. 
The trajectory is assumed to be $C^9$ in the closed interval 
$[\tau_0, \tau]$. 
We wish to evaluate the limit $\epsilon\to0$ of 
the transition rate given by 
(\ref{defexcitation-sharp}) with the 
correlation function~(\ref{corrSchlicht}). 

Writing $\Delta\tau := \tau - \tau_0$, the regularised transition rate
(\ref{defexcitation-sharp}) reads 
\begin{equation}
\label{Fsch}
\dot{F}_{\tau} (\omega)=\frac{1}{2\pi^2}\ 
\mathrm{Re}
\int_{0}^{\Delta\tau}
\mathrm{d}s\,
\frac{\mathrm{e}^{-i\omega s}}
{\bigl( \mathsf{x}-\mathsf{x}'
-i\epsilon 
(\dot{\mathsf{x}}+\dot{\mathsf{x}}') 
\bigr)^2}
\ , 
\end{equation}
where we have suppressed the index~$\epsilon$ on the left-hand side. 
Decomposing 
$\dot{F}_{\tau} (\omega)$ into its even and odd parts in $\omega$
as 
$\dot{F}_{\tau} (\omega) = 
\dot{F}^{\mathrm{even}} _{\tau} (\omega)
+ 
\dot{F}^{\mathrm{odd}}_{\tau} (\omega)$, 
we find
\begin{subequations}
\label{eq:even-and-odd}
\begin{align}
\label{eq:even}
\dot{F}^{\mathrm{even}}_{\tau} (\omega)
&= 
\frac{1}{2\pi^2}
\int_{0}^{\Delta\tau}
\mathrm{d}s 
\, 
\frac{
\bigl[{(\Delta\mathsf{x})}^2 -\epsilon^2\vsigma^2\bigr]
\cos(\omega s)}
{{\bigl[\epsilon^2\vsigma^2-{(\Delta\mathsf{x})}^2\bigr]}^2
+4\epsilon^2 {(\vsigma\cdot\Delta \mathsf{x})}^2}
\ ,  
\\
\label{eq:odd}
\dot{F}^{\mathrm{odd}}_{\tau} (\omega)
&= 
\frac{1}{\pi^2}
\int_{0}^{\Delta\tau}
\mathrm{d}s 
\, 
\frac{\epsilon(\vsigma\cdot \Delta
\mathsf{x})\sin(\omega s)}
{{\bigl[\epsilon^2\vsigma^2-{(\Delta\mathsf{x})}^2\bigr]}^2
+4\epsilon^2 {(\vsigma\cdot\Delta \mathsf{x})}^2}
\ , 
\end{align}
\end{subequations}
where 
$\Delta \mathsf{x} := \mathsf{x}(\tau) - \mathsf{x}(\tau- s)$, 
${(\Delta \mathsf{x})}^2 := 
(\Delta \mathsf{x}) \cdot (\Delta \mathsf{x})$ and 
$\vsigma := 
\dot{\mathsf{x}}(\tau)+\dot{\mathsf{x}}(\tau-s)$. 
We further decompose 
$\dot{F}^{\mathrm{even}}
= 
\dot{F}_{<}^{\mathrm{even}}
+ 
\dot{F}_{>}^{\mathrm{even}}$ 
and 
$\dot{F}^{\mathrm{odd}}
= 
\dot{F}_{<}^{\mathrm{odd}}
+ 
\dot{F}_{>}^{\mathrm{odd}}$, 
where the subscripts $<$ and $>$ 
refer respectively 
to the integration subintervals 
$s \in \bigl[0,\sqrt{\epsilon}\bigr]$ 
and 
$s \in \bigl[\sqrt{\epsilon}, \Delta\tau\bigr]$ 
and we have suppressed $\tau$ and~$\omega$. 

We consider each of the four terms in turn. 
To begin we note the small $s$ expansions 
\begin{subequations}
\label{eq:small-s}
\begin{align}
\label{eq:small-s-Deltax}
{(\Delta\mathsf{x})}^2
&=
-s^2
-\tfrac{1}{12}
{\ddot{\mathsf{x}}}^2
s^4
+O(s^5)
\ , 
\\
\vsigma\cdot\Delta \mathsf{x}
&=
-2s
- 
\tfrac{1}{3}
{\ddot{\mathsf{x}}}^2
s^3
+O(s^4)
\ , 
\\
\vsigma^2
&=
-4
-
{\ddot{\mathsf{x}}}^2
s^2
+O(s^3)
\ . 
\end{align}
\end{subequations}

Consider $\dot{F}_{>}^{\mathrm{even}}$ and $\dot{F}_{>}^{\mathrm{odd}}$. 
For fixed~$s$, the integrand in 
(\ref{eq:even}) tends to $\cos(\omega s) / {(\Delta\mathsf{x})}^2$. 
If we make this replacement under the integral, the resulting error in
$\dot{F}_{>}^{\mathrm{even}}$ can be arranged into the form 
\begin{equation}
\label{even}
\frac{1}{2\pi^2} 
\int_{\sqrt{\epsilon}}^{\Delta\tau}\mathrm{d}s 
\cos(\omega s)\frac{\epsilon^2}{{\Bigl[{(\Delta\mathsf{x})}^2\Bigr]}^2}
\frac{
\displaystyle
\vsigma^2
-4\frac{{(\vsigma\cdot\Delta \mathsf{x})}^2}{{(\Delta\mathsf{x})}^2}
-\epsilon^2 \frac{{(\vsigma^2)}^2}{{(\Delta\mathsf{x})}^2}}
{
\displaystyle
\left\{
\left( 1-\epsilon^2\frac{\vsigma^2}{{(\Delta\mathsf{x})}^2}\right)^2
+4\epsilon^2\frac{{(\vsigma\cdot\Delta \mathsf{x})}^2}
{{\Bigl[{(\Delta\mathsf{x})}^2\Bigr]}^2}\right\} }
\ . 
\end{equation}
From (\ref{eq:small-s}) it follows that $\vsigma^2$,
${(\vsigma\cdot\Delta \mathsf{x})}^2 / {(\Delta\mathsf{x})}^2$ and
$\epsilon/{(\Delta\mathsf{x})}^2$ are all bounded in absolute value
over the interval of integration by constants independent of
$\epsilon$ as $\epsilon \to 0$, uniformly in~$s$. Since $\bigl\vert
{(\Delta\mathsf{x})}^2 \bigr\vert \ge s^2$, the absolute value of the
integrand in (\ref{even}) is thus bounded by a constant times
$\epsilon^2/s^4$ and the integral is of order $O(\sqrt{\epsilon})$. A
similar estimate for the integrand in (\ref{eq:odd}) shows that
$\dot{F}^{\mathrm{odd}}_{>} = O(\sqrt{\epsilon})$. Hence
\begin{equation}
\label{eq:large-even-and-odd}
\dot{F}^{\mathrm{even}}_{>} 
+ 
\dot{F}^{\mathrm{odd}}_{>} 
= 
\frac{1}{2\pi^2}
\int_{\sqrt{\epsilon}}^{\Delta\tau}
\mathrm{d}s 
\, 
\frac{
\cos(\omega s)}
{{(\Delta\mathsf{x})}^2}
\ \ \ 
+ O(\sqrt{\epsilon})
\ . 
\end{equation}

Consider then $\dot{F}_{<}^{\mathrm{even}}$ 
and $\dot{F}_{<}^{\mathrm{odd}}$. 
At $s=0$, the denominator in the integrands in (\ref{eq:even-and-odd}) 
is of order~$\epsilon^4$. To control the denominator for 
$s \in \bigl[0,\sqrt{\epsilon}\bigr]$, we expand 
${(\Delta\mathsf{x})}^2$ in $s$ to order~$s^8$,
$\vsigma\cdot\Delta \mathsf{x}$ 
to order $s^5$ and $\vsigma^2$ to order $s^4$: 
this gives the denominator accurately to order~$\epsilon^5$. 
Writing $s = \sqrt{\epsilon} r$, where $0\le r \le1$, we find 
\begin{equation}
\label{eq:den-finalexpansion}
{\bigl[\epsilon^2\vsigma^2-{(\Delta\mathsf{x})}^2\bigr]}^2
+4\epsilon^2 {(\vsigma\cdot\Delta \mathsf{x})}^2
= 
\epsilon^2 {(4\epsilon+r^2)}^2
\left[ 
1 + \tfrac16  {\ddot{\mathsf{x}}}^2
\epsilon r^2 
+ O(\epsilon^{3/2}) 
\right] 
\ , 
\end{equation}
where taking out the factor 
${(4\epsilon+r^2)}^2$ 
has allowed all the terms that depend on the higher 
derivatives of $\mathsf{x}$ to be grouped into the 
$O(\epsilon^{3/2})$ term, uniformly in~$r$. 

In $\dot{F}_{<}^{\mathrm{odd}}$, it suffices to 
keep in the denominator just the leading term in 
(\ref{eq:den-finalexpansion}) and in the numerator 
just the leading power of~$s$. 
We find 
\begin{align}
\dot{F}^{\mathrm{odd}}_<
&= 
- \frac{2\omega \sqrt{\epsilon}}{\pi^2}
\int_{0}^{1}
\mathrm{d}r 
\, 
\frac{r^2}
{{(4\epsilon+r^2)}^2}
\bigl[1 + O(\epsilon) \bigr]
\notag
\\
&= 
- \frac{\omega}{4\pi}
+ O(\sqrt{\epsilon})
\ , 
\label{eq:small-odd}
\end{align}
where the integration is elementary. 
In $\dot{F}_{<}^{\mathrm{even}}$, 
we use (\ref{eq:den-finalexpansion})
in the denominator 
and expand the numerator to next-to-leading order in~$s$. 
We find 
\begin{align}
\dot{F}^{\mathrm{even}}_<
&= 
\frac{1}{2 \pi^2 \sqrt{\epsilon}}
\int_{0}^{1}
\mathrm{d}r 
\, 
\frac{(4\epsilon - r^2)
\left[
1 
- \left(
\tfrac{1}{12}{\ddot{\mathsf{x}}}^2
+ \tfrac12 \omega^2
\right)
\epsilon r^2 
\right]
+ r^2 O(\epsilon^{3/2}) 
+ O(\epsilon^{5/2})}
{{(4\epsilon+r^2)}^2}
\notag
\\
&=
\frac{1}{2 \pi^2 \sqrt{\epsilon}}
+ O(\sqrt{\epsilon}) 
\ , 
\label{eq:small-even}
\end{align}
where the integrations are elementary. 

Combining (\ref{eq:large-even-and-odd}), 
(\ref{eq:small-odd}) and 
(\ref{eq:small-even}),
and writing 
$\epsilon^{-1/2} = {(\Delta\tau)}^{-1} 
+ \int_{\sqrt{\epsilon}}^{\Delta\tau} 
s^{-2} \, \mathrm{d}s$, we obtain 
\begin{equation}
\label{eq:lorentzian-almost}
\dot{F}_{\tau}(\omega)
=
-\frac{\omega}{4\pi}+\frac{1}{2\pi^2}
\int_{\sqrt{\epsilon}}^{\Delta\tau}\textrm{d}s
\left( 
\frac{\cos (\omega s)}{{(\Delta \mathsf{x})}^2} 
+ 
\frac{1}{s^2} 
\right) 
\ \ +\frac{1}{2\pi^2 \Delta \tau}
\ \ +O(\sqrt{\epsilon})
\ . 
\end{equation}
As it follows from (\ref{eq:small-s-Deltax}) 
that the integrand in 
(\ref{eq:lorentzian-almost})
has a small $s$ expansion that starts with a constant term, 
taking the limit $\epsilon\to0$ yields
\begin{equation}
\label{resultado1}
\dot{F}_{\tau}(\omega)
=
-\frac{\omega}{4\pi}+\frac{1}{2\pi^2}
\int_0^{\Delta\tau}\textrm{d}s
\left( 
\frac{\cos (\omega s)}{{(\Delta \mathsf{x})}^2} 
+ 
\frac{1}{s^2} 
\right) 
\ \ +\frac{1}{2\pi^2 \Delta \tau}
\ . 
\end{equation}
Formula (\ref{resultado1}) is the promised result.

\subsection{Switch-on in the asymptotic past} 
\label{subsec:lor-as}

We now turn to a detector that is switched on in the asymptotic
past. Two qualitatively different situations can arise here. One
occurs for trajectories that are defined for arbitrarily negative
proper times, the other for trajectories that come from infinity
within finite proper time. 

We note first that formula (\ref{resultado1}) has in both situations
a well-defined limit. In the former case $\tau_0$ is
replaced by $-\infty$ and we obtain
\begin{equation}
\label{resultado1-infty}
\dot{F}_{\tau}(\omega)
=
-\frac{\omega}{4\pi}+\frac{1}{2\pi^2}
\int_0^{\infty}\textrm{d}s
\left( 
\frac{\cos (\omega s)}{{(\Delta \mathsf{x})}^2} 
+ 
\frac{1}{s^2} 
\right) 
\ . 
\end{equation}
In the latter case $\tau_0$ in (\ref{resultado1}) is understood as the
(asymptotic) value of the proper time at which the trajectory starts
out at infinity. In either situation the integral is convergent in
absolute value since 
$s^2 \leq \left\vert {(\Delta \mathsf{x})}^2 \right\vert$. 

The only step that changes 
in the analysis of subsection 
(\ref{subsec:lor-sharp})
is that the estimates for 
$\dot{F}^{\mathrm{even}}_{>}$ and 
$\dot{F}^{\mathrm{odd}}_{>}$ need to control the 
integrand also 
as $s$ approaches respectively
$\infty$ or $\tau -\tau_0$. 
Inspection of the integrand in 
(\ref{even}) 
and the similar arrangement of the integrand in 
$\dot{F}^{\mathrm{odd}}_{>}$
shows that a 
sufficient condition is to assume that 
the quantities 
\begin{equation} 
\label{eq:infty-bounds-lor}
\frac{\vsigma^2}{{(\Delta \mathsf{x})}^2}
\ \ , \ \ 
\frac{\vsigma\cdot\Delta \mathsf{x}}{{(\Delta \mathsf{x})}^2}
\end{equation}
remain bounded as $s$ approaches these asymptotic values. 
Taking the 
limit $\epsilon\to0$ then yields 
result (\ref{resultado1-infty}) in the former 
case and result (\ref{resultado1}) in the latter case. 

We shall show in section
\ref{sec:applications} that 
quantities (\ref{eq:infty-bounds-lor}) do remain bounded as
$s\to\infty$ for a number of interesting trajectories that are
defined for arbitrarily negative proper times, including in particular
all stationary trajectories.

\subsection{Summary}

To summarise, we have obtained an explicit expression for the
excitation rate of a particle detector regulated with the Lorentzian
spatial profile once the zero-size limit has been taken and the
regulator has disappeared. For detectors switched on at finite
$\tau_0$ the formula is (\ref{resultado1}) and is valid for arbitrary
trajectories; for detectors switched on at the asymptotic past the
formula is (\ref{resultado1-infty}) (or again (\ref{resultado1}) in
the special case of motion coming from infinity in a finite
$\Delta\tau$) and is valid if quantities (\ref{eq:infty-bounds-lor})
are bounded over the trajectory.

Note that formulas (\ref{resultado1}) and (\ref{resultado1-infty})
hold both for $\omega>0$ and $\omega<0$. We shall return to this
point in subsection~\ref{subsec:falloff-etc}.

In the following section we show that the same results can be obtained from
more general spatial profiles, while in Appendix A we show that a
similar calculation using the standard $i\epsilon$ regularisation
gives the same result with an added Lorentz-noninvariant term.

\section{Spatially extended detector model with a general profile}
\label{sec:generalprofile}

In this section we examine a spatially extended detector model that is
motivated by the spatially smeared detector introduced in
section~\ref{sec:detectormodels}. Although we will not be able to
establish a precise connection between this model and the spatially
smeared field, the interest of the model is that it does capture some
of the contributions from spatial smearing and these contributions
will be seen to be independent of the profile function,
yielding the
same zero-size limit as in 
section~\ref{sec:lor-limit}. 

We again address first a detector switched on at a finite proper time
and then a detector switched on in the asymptotic past.

\subsection{Sharp switch-on} 

We again denote by 
$\tau_0$ the moment of the sharp switch-on
and by $\tau$ the moment of observation, assuming $\tau> \tau_0$. 
The trajectory is now assumed to be real analytic in the closed interval 
$[\tau_0, \tau]$. 

Consider the spatially smeared detector model of section
section~\ref{sec:detectormodels}, and assume that the function $f$ in
the profile (\ref{eq:profile-scaling}) is smooth and has compact
support. Note that $f$ is not assumed to be spherically symmetric. 
Note also that 
our discussion will not cover 
the Lorentzian profile~(\ref{lorentzian}), 
which is not of compact support.

Substituting the 
smeared field operator 
(\ref{smeared}) in (\ref{defexcitation-sharp}) 
and formally interchanging the integrals, 
we obtain for the transition rate the formula 
\begin{equation}
\label{eq:smearedFdot}
\dot{F}_{\tau}^{(\epsilon)} (\omega)
=
\int 
\mathrm{d^3}\xi
\,
\mathrm{d^3}\xi' 
\; 
f_{\epsilon}(\boldsymbol{\xi}) 
\, 
f_{\epsilon}(\boldsymbol{\xi}')
\, 
{G}_{\tau,\tau_0} 
(\boldsymbol{\xi} , \boldsymbol{\xi}'; \omega)
\ , 
\end{equation}
where 
\begin{equation}
\label{eq:G-def}
G_{\tau,\tau_0} 
(\boldsymbol{\xi} , \boldsymbol{\xi}' ; \omega)
:=  
2 \,
\mathrm{Re}\int_{0}^{\Delta\tau} \mathrm{d}s
\,\,
\mathrm{e}^{-i \omega s}
\, 
\langle 0\vert 
\phi \bigl(\mathsf{x} (\tau,\boldsymbol{\xi}) \bigr)
\phi \bigl(\mathsf{x}(\tau-s,\boldsymbol{\xi}') \bigr)
\vert 0\rangle
\end{equation}
and $\Delta\tau := \tau - \tau_0$. 
The difficulty with 
(\ref{eq:smearedFdot})
and 
(\ref{eq:G-def})
is that the latter formula suffers at 
$\boldsymbol{\xi} = \boldsymbol{\xi}'$ from the same ambiguity as the
unsmeared transition rate~(\ref{defexcitation-sharp}). 
However, we shall show in subsection \ref{subsec:G} 
that ${G}_{\tau,\tau_0}$ 
is pointwise well defined by (\ref{eq:G-def})
whenever 
$\vert\boldsymbol{\xi}\vert$
and 
$\vert\boldsymbol{\xi}'\vert$ 
are sufficiently small and $\boldsymbol{\xi} \ne \boldsymbol{\xi}'$.
We shall further show that the integral in 
(\ref{eq:smearedFdot}) over the subset  
$\boldsymbol{\xi} \ne \boldsymbol{\xi}'$ 
is well defined for sufficiently small~$\epsilon$. 

Motivated by these observations, we take the instantaneous transition
rate in our smeared detector model to be
\emph{defined\/} for sufficiently small $\epsilon$ by \begin{equation}
\label{eq:smearedFdot-def}
\dot{F}_{\tau}^{(\epsilon)} (\omega)
: =
\int_{\boldsymbol{\xi} \ne \boldsymbol{\xi}'}
\mathrm{d^3}\xi
\,
\mathrm{d^3}\xi' 
\; 
f_{\epsilon}(\boldsymbol{\xi}) 
\, 
f_{\epsilon}(\boldsymbol{\xi}')
\, 
{G}_{\tau,\tau_0} 
(\boldsymbol{\xi} , \boldsymbol{\xi}'; \omega)
\ . 
\end{equation}
We shall show in 
subsection \ref{subsec:G} 
that the limit of $\dot{F}_{\tau}^{(\epsilon)} (\omega)$ 
(\ref{eq:smearedFdot-def})
as $\epsilon \to 0$ exists, 
is indepedent of the profile function and 
given by~(\ref{resultado1}). 

If the passage from (\ref{defexcitation-sharp}) with (\ref{smeared})
to (\ref{eq:smearedFdot-def}) can be justified in a sense in which
${G}_{\tau,\tau_0}$ does not contain a distribution with support at
$\boldsymbol{\xi} =\boldsymbol{\xi}'$, our model is equivalent to
spatial smearing with a profile of compact support. As our model
yields the same transition rate as spatial smearing with the
Lorentzian profile function (which is not of compact support), we are
led to suspect that the equivalence of our model to spatial smearing
could be established for at least some classes of profile
functions. We shall not pursue this question further in this paper.

Readers who wish to skip the technical estimates on
${G}_{\tau,\tau_0}$ may prefer to proceed directly to
subsection~\ref{subsec:taunought-infty}.

\subsection{Estimates for ${G}_{\tau,\tau_0}$}
\label{subsec:G}

As the profile function 
(\ref{eq:profile-scaling}) has by assumption compact support, 
it suffices in (\ref{eq:smearedFdot-def}) to define and examine 
${G}_{\tau,\tau_0}$ 
(\ref{eq:G-def}) 
at small 
$\vert\boldsymbol{\xi}\vert$
and 
$\vert\boldsymbol{\xi}'\vert$. 
To control the smallness, we introduce a positive parameter $\delta$ 
and assume 
$\vert\boldsymbol{\xi}\vert < \delta$, 
$\vert\boldsymbol{\xi}'\vert < \delta$ and 
$\boldsymbol{\xi} \ne \boldsymbol{\xi}'$. 
The limit of interest is $\delta \to 0$, where 
$\tau$, $\tau_0$ and $\omega$ are regarded as fixed. 

As the singularity of the Wightman function is on the light cone, the
integral over $s$ in (\ref{eq:G-def}) has a singularity precisely when
the vector
\begin{equation}
\label{Hmu}
\mathsf{H} 
:=
\mathsf{x}(\tau)-\mathsf{x}(\tau-s)
+\xi^{i} \mathsf{e}_i(\tau)
- \xi'^{i} \mathsf{e}_i(\tau-s) 
\end{equation}
is null. 
For sufficiently small~$\delta$, 
it follows from the construction of the 
Fermi-Walker coordinates \cite{mtw} 
that $\mathsf{H}$ is spacelike at $s=0$, 
future timelike at $s=\Delta\tau$ and null at exactly 
one intermediate value of~$s$, 
which we denote by~$s^*$. 
This means that we can define the integral over $s$ 
by representing the Wightman function as in~(\ref{dist}): 
Decomposing $\mathsf{H}$ into its temporal and spatial components as 
$\mathsf{H} =: (H^0,\mathbf{H})$, we obtain 
\begin{align}
&
\langle 0\vert 
\phi \bigl(\mathsf{x} (\tau,\boldsymbol{\xi}) \bigr)
\phi \bigl(\mathsf{x}(\tau-s,\boldsymbol{\xi}') \bigr)
\vert 0\rangle
\notag
\\
\noalign{\medskip}
& 
\ \ 
= 
\frac{1}{8\pi^2}\frac{1}{\vert\mathbf{H}\vert}
\left[ P\left(
\frac{1}{\vert\mathbf{H}\vert-H^0}\right)
+P\left( \frac{1}{\vert\mathbf{H}\vert+H^0}\right)
-i\pi \delta(\vert\mathbf{H}\vert-H^0)
+ i\pi \delta(\vert\mathbf{H}\vert+H^0) \right] 
\ , 
\label{dist2}
\end{align}
which contains a prescription for 
integrating over $s=s^*$. 
Note that since $\mathsf{H}$ is nonvanishing for all~$s$, the 
integral over any zeroes of $\mathbf{H}$ is nonsingular 
despite the overall factor~$1/\vert\mathbf{H}\vert$. 
Note also that this is the step where we need the 
assumption 
$\boldsymbol{\xi} \ne \boldsymbol{\xi}'$. 
When $\boldsymbol{\xi} = \boldsymbol{\xi}'$, 
$\mathsf{H}$ vanishes at $s=0$ 
and the integral over $s$ faces the same ambiguity at $s\to0$ 
as the unsmeared integral~(\ref{defexcitation-sharp}). 

Since $\vert\mathbf{H}\vert+H^0>0$, the term proportional to
$\delta(\vert\mathbf{H}\vert+H^0)$ 
in (\ref{dist2})
gives a vanishing contribution to the integral, and in
the term involving $P\bigl[ 1/(\vert\mathbf{H}\vert+H^0)\bigr]$ 
the principal value symbol
is redundant and can be dropped. The contribution from the remaining
principal value term can be converted into a contour integral by the
identity
\begin{equation}
\int_{0}^{\Delta\tau}\mathrm{d}s\,\, 
P \bigl( g(s) \bigr) 
=
-i\pi\mathrm{Res}
\bigl( g(s) \bigr)_{s^*} 
+\int_{C}\mathrm{d}s \, g(s)
\ , 
\end{equation}
where the contour $C$ circumvents the pole at $s=s^*$ in the lower
half of the complex $s$ plane. The contribution from the residue will
then cancel the contribution from the remaining delta-function, 
since $\mathrm{d}(\vert\mathbf{H}\vert-H^0)/\mathrm{d}s$
is negative at $s=s^*$: 
This is because $\mathsf{H}$ is spacelike for 
$s< s^*$ and \emph{future\/} timelike for $s> s^*$. 
We thus obtain
\begin{equation}
\label{eq:GfromH}
G_{\tau,\tau_0} 
(\boldsymbol{\xi} , \boldsymbol{\xi}' ; \omega)
= 
\frac{1}{2\pi^2}\mathrm{Re}\int_C \mathrm{d}s
\, 
\frac{\mathrm{e}^{-i\omega s}}{{\mathsf{H}}^2(s,\tau)}
\ , 
\end{equation}
where the dependence of $\mathsf{H}$ on 
$\boldsymbol{\xi}$ and~$\boldsymbol{\xi}'$ has been suppressed. 

We wish to compute 
${G}_{\tau,\tau_0}$ 
from 
(\ref{eq:GfromH}) 
in the limit $\delta\to0$. 
The technical subtlety in this computation
is that 
although $s^*$ is positive, 
it may be arbitrarily small compared with~$\delta$. 

We first establish some facts about the 
small $s$ behaviour of~${\mathsf{H}}^2$. 
From~(\ref{Hmu}), we obtain 
\begin{equation}
\label{eq:H2-split}
{\mathsf{H}}^2
 = 
(\Delta \mathsf{x})^2 +
2 \, 
\Delta \mathsf{x}
\cdot 
(\xi^i \mathsf{e}_i - 
\xi'^j \mathsf{e}'_j)
+ (\xi^i \mathsf{e}_i - 
\xi'^j \mathsf{e}'_j)^2
\ , 
\end{equation}
where 
\begin{subequations}
\begin{align}
\Delta \mathsf{x}
& := 
\mathsf{x}(\tau) - \mathsf{x}(\tau - s)
\ , 
\\
\mathsf{e}_i 
& 
:= 
\mathsf{e}_i (\tau) 
\ , 
\\
\mathsf{e}'_i 
& 
:= 
\mathsf{e}_i (\tau - s) 
\ . 
\end{align}
\end{subequations}
Note that 
\begin{equation}
(\xi^i \mathsf{e}_i - 
\xi'^j \mathsf{e}'_j)^2
= 
{\vert \boldsymbol{\xi} - \boldsymbol{\xi}' \vert}^2
+ 2 
\xi^i \xi'^j 
\left( \delta_{ij} - 
\mathsf{e}_i \cdot \mathsf{e}'_j
\right)
\ . 
\end{equation}
Expanding $\mathsf{x}(\tau-s)$ and $\mathsf{e}(\tau-s)$ 
in powers of $s$ gives the small $s$ expansions 
(\ref{eq:small-s-Deltax}) 
and 
\begin{equation}\label{eq:SA-small}
\mathsf{H}^2 = s_0^2 + \sum_{j=2}^{\infty} H_j s^j
\end{equation}
where $s_0 := \vert \boldsymbol{\xi} - \boldsymbol{\xi}' \vert$
and 
\begin{align}
H_2 
& = 
-1 - ( \xi^i + \xi'^i ) 
(\ddot{\mathsf{x}} \cdot \mathsf{e}_i)
- \xi^i \xi'^j 
(\ddot{\mathsf{x}} \cdot \mathsf{e}_i)
(\ddot{\mathsf{x}} \cdot \mathsf{e}_j) 
= -1 + O(\delta)
\ , 
\notag 
\\
H_3 
& = \tfrac13
( \xi^j + 2\xi'^j )  
(\dddot{\mathsf{x}} \cdot \mathsf{e}_j) +
\tfrac13 
\xi^i \xi'^j 
\bigl[ 
(\dddot{\mathsf{x}} \cdot \mathsf{e}_i) 
(\ddot{\mathsf{x}} \cdot \mathsf{e}_j) 
+ 2 
(\ddot{\mathsf{x}} \cdot \mathsf{e}_i) 
(\dddot{\mathsf{x}} \cdot \mathsf{e}_j) 
\bigr] 
= 
O(\delta)
\ , 
\notag
\\
H_j 
& = 
O(\delta^0) \ \ \textrm{for} \ \ j\ge4
\ .
\label{eq:SA-small-coeffs}
\end{align}
We have here used the normalisation condition ${\dot{\mathsf{x}}}^2=-1$
and its consequences for the higher derivatives, 
the Fermi-Walker transport equation for the tetrad and the bounds 
$\vert\boldsymbol{\xi}\vert < \delta$ 
and 
$\vert\boldsymbol{\xi}'\vert < \delta$. 

\begin{figure}[t]
\centering
\includegraphics[width=0.8\textwidth]{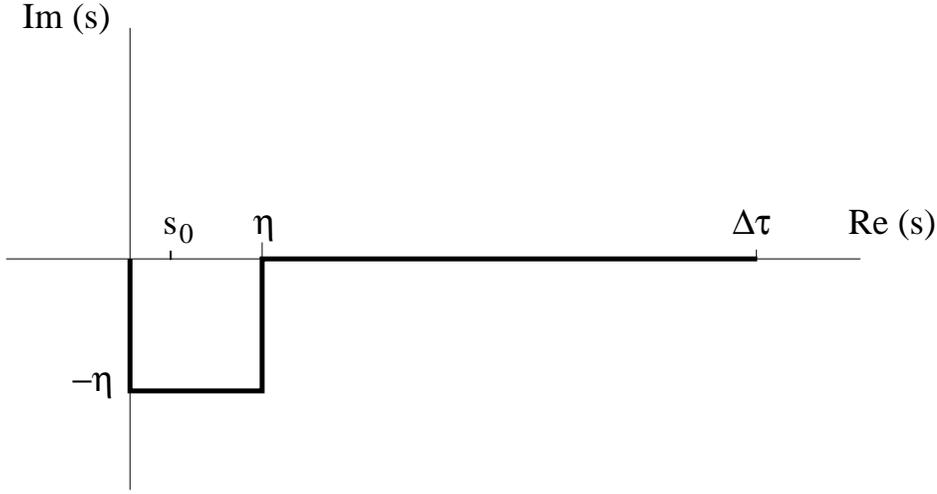}
\caption{The contour $C$ (\ref{eq:contour}) in the complex $s$ plane.
We have written $\auxeps:= \sqrt{\delta}$.}
\label{fig:contour}
\end{figure}

Next, we specify the contour. 
For sufficiently small~$\delta$, 
(\ref{eq:SA-small}) and (\ref{eq:SA-small-coeffs})
show that 
$\mathsf{H}$ is timelike at $s=\sqrt{\delta}$, 
from which it follows that $s^* < \sqrt{\delta}$. 
We may thus take the contour $C$ to consist 
of four straight lines
$C_i$, $i = 1,2,3,4$, as shown in Figure~\ref{fig:contour}: 
\begin{align}
& 
C_1 : \ 
s= -i r, \ \ 0 \le r \le \sqrt{\delta}
\ , 
\notag
\\
& 
C_2 : \ 
s= -i \sqrt{\delta} + r, \ \ 0 \le r \le \sqrt{\delta}
\ , 
\notag
\\
& 
C_3 : \ 
s= \sqrt{\delta} (1 - i) + i r  , \ \ 0 \le r \le \sqrt{\delta}
\ , 
\notag
\\
& 
C_4 : \ 
\sqrt{\delta} \le s \le \Delta\tau
\ . 
\label{eq:contour} 
\end{align}
Note that since $s_0 < 2\delta$, 
we have $s_0 < \sqrt{\delta}$ for sufficiently small~$\delta$.

We show in appendix 
\ref{app:contourestimates}
that the contribution 
to 
${G}_{\tau,\tau_0}$ 
from
${C_2 \cup C_3 \cup C_4}$ reads 
\begin{equation}
\label{eq:GSA-C2C3C4:2}
{G}_{\tau,\tau_0}^{C_2 \cup C_3 \cup C_4} 
(\boldsymbol{\xi} , \boldsymbol{\xi}' ; \omega)
= 
-\frac{\omega}{4\pi}
+ 
\frac{1}{2\pi^2}\int_0^{\Delta \tau} 
\mathrm{d}s
\, 
\left(
\frac{\cos (\omega s)}{{(\Delta \mathsf{x})}^2} 
+  \frac{1}{s^2}
\right)
\ \ +\frac{1}{2\pi^2 \Delta \tau}
\ + O\bigl(\sqrt{\delta} \, \bigr)
\ . 
\end{equation}
It follows that 
${G}_{\tau,\tau_0}^{C_2 \cup C_3 \cup C_4} 
(\boldsymbol{\xi} , \boldsymbol{\xi}' ; \omega)$ 
has a limit as 
$(\boldsymbol{\xi}, \boldsymbol{\xi}') \to 
(\boldsymbol{0}, \boldsymbol{0})$ with 
$\boldsymbol{\xi} \ne \boldsymbol{\xi}'$, 
given by dropping the 
$O$-term from~(\ref{eq:GSA-C2C3C4:2}). 
We also show in appendix 
\ref{app:contourestimates}
that the contribution
to 
${G}_{\tau,\tau_0}$ 
from
${C_1}$
is of the form 
\begin{equation}
\label{eq:GSA-C1:2}
{G}_{\tau,\tau_0}^{C_1} 
(\boldsymbol{\xi} , \boldsymbol{\xi}' ; \omega)
= 
O(\delta) \ln(s_0)
+ O\bigl(\sqrt{\delta} \, \bigr)
\ . 
\end{equation}
The logarithmic term in (\ref{eq:GSA-C1:2}) implies that 
${G}_{\tau,\tau_0}^{C_1} 
(\boldsymbol{\xi} , \boldsymbol{\xi}' ; \omega)$ 
does not have a limit as 
$(\boldsymbol{\xi}, \boldsymbol{\xi}') \to 
(\boldsymbol{0}, \boldsymbol{0})$ with 
$\boldsymbol{\xi} \ne \boldsymbol{\xi}'$, 
but the coefficient of this term and the 
integrability of the logarithm 
imply that 
the contribution 
of ${G}_{\tau,\tau_0}^{C_1}$
to 
the transition rate (\ref{eq:smearedFdot-def}) vanishes in the limit 
$\epsilon \to0$. 
These estimates imply that the 
$\epsilon \to0$ limit 
of the transition rate (\ref{eq:smearedFdot-def}) 
exists and is given by~(\ref{resultado1}).

\subsection{Switch-on in the asymptotic past}
\label{subsec:taunought-infty}

For a detector switched on in the asymptotic past, we must again
consider the situation in which the trajectory is defined for
arbitrarily negative proper times and the situation in which the
trajectory comes from infinity within finite proper time. We shall
show in appendix
\ref{app:contourestimates} that in either case the analysis of 
subsection \ref{subsec:G} generalises if the trajectory 
is sufficiently well-behaved in the asymptotic 
past. A~sufficient condition is that the quantities 
\begin{equation} 
\label{eq:infty-bounds}
\frac{\mathsf{e}_i \cdot\mathsf{e}'_j}{{(\Delta \mathsf{x})}^2}
\ \ , \ \ 
\frac{\Delta\mathsf{x}\cdot\mathsf{e}_i}{{(\Delta \mathsf{x})}^2}
\ \ , \ \ 
\frac{\Delta\mathsf{x}\cdot\mathsf{e}'_i}{{(\Delta \mathsf{x})}^2}
\ , 
\end{equation}
all remain bounded as $s$ grows to its asymptotic value, be it
$+\infty$ or~$\Delta\tau$. We shall show 
in section \ref{sec:applications}
that this condition,
just as the equivalent one for~(\ref{eq:infty-bounds-lor}), is
satisfied for many trajectories of interest and in particular for all
stationary ones, whose associated particle spectra can therefore be
calculated from~(\ref{resultado1-infty}).

\section{Applications}
\label{sec:applications}

In this section we shall first discuss some 
general properties of our transition rate formulas 
(\ref{resultado1}) 
and 
(\ref{resultado1-infty}) 
and then apply these formulas to specific trajectories of interest.

\subsection{Causality, parity and falloff} 
\label{subsec:falloff-etc}

We have already noted that the integral formulas 
(\ref{resultado1}) 
and 
(\ref{resultado1-infty}) 
have 
the technical advantage of no longer 
containing a regularisation parameter 
that would have to be taken to zero after integration. 
The formulas also show explicitly that the response is causal, 
as emphasised by Schlicht
\cite{schlicht} in the context of formulas 
(\ref{defexcitation-sharp-infty}) and~(\ref{corrSchlicht}): 
The transition rate at time $\tau$ is a 
functional of the detector's trajectory at times prior to~$\tau$. 

Formulas (\ref{resultado1}) and (\ref{resultado1-infty}) give the
spectrum as decomposed into its odd and even parts in~$\omega$. The
odd part is always equal to $-\omega/4\pi$, and only the even part is
affected by the past of the trajectory. Departures from inertial
motion have thus an equal probability of inciting upward and
downward transitions in the detector.

Another useful decomposition of the spectrum is into the inertial part
and the non-inertial correction, as introduced 
for stationary trajectories in \cite{letaw-pfautsch,letaw}. 
This may be obtained by adding and
subtracting $\cos(\omega s) / s^2$ under the integrals in
(\ref{resultado1}) and~(\ref{resultado1-infty}). For a detector
switched on in the asymptotic past of infinite proper time,
(\ref{resultado1-infty}) gives
\begin{align}
{\dot F}_{\tau}(\omega)
& =
-\frac{\omega}{4\pi}
+\frac{1}{2\pi^2}
\int_0^{\infty}\textrm{d}s
\left[ 
\frac{\cos (\omega s)}{{(\Delta \mathsf{x})}^2} 
+\frac{\cos(\omega s)}{s^2}
+ \left( 
\frac{1}{s^2}-\frac{\cos(\omega s)}{s^2}
\right) 
\right] 
\notag
\\
& = 
-\frac{\omega}{2\pi}\Theta(-\omega)
+\frac{1}{2\pi^2}\int_0^{\infty}\textrm{d}s\cos (\omega s)
\left( 
\frac{1}{{(\Delta \mathsf{x})}^2} 
+\frac{1}{s^2}
\right) 
\ , 
\label{inermasacc}
\end{align}
where the last expression is obtained by the method of residues. The
first term in (\ref{inermasacc}) is the spectrum in
inertial motion and the integral term is the 
correction due to acceleration. 
Note that the correction is nonvanishing for 
\emph{every\/} noninertial trajectory. 
Note also that the 
transition rate completely determines 
${(\Delta \mathsf{x})}^2$ 
as a function of $\tau$ and~$s$: 
as the inertial term vanishes for positive~$\omega$, 
inverting the cosine transform in (\ref{inermasacc})
gives  
\begin{equation}
\frac{1}{{(\Delta \mathsf{x})}^2} 
+\frac{1}{s^2}
=
4\pi\int_0^{\infty}\mathrm{d}\omega 
\, \dot{F}_{\tau}(\omega)\cos(\omega s)
\ . 
\end{equation}

We can use (\ref{inermasacc}) to obtain another interesting property
of the spectrum. For sufficiently differentiable trajectories, one expects 
the non-inertial effects to become negligible at small length
scales and the spectrum thus to approach the inertial spectrum at high
frequencies. To verify that this indeed happens, we use the following
theorem~\cite{wong}: If the function $h$ is $C^\infty$ in $[a,\infty)$
and $h^{(n)}(s)=O(s^{-1-\epsilon})$ as $s\rightarrow\infty$ for some
$\epsilon>0$ and every $n\geq0$, then
\begin{equation}
\label{theorem}
\int_a^\infty\mathrm{d}s \, h(s)
\mathrm{e}^{ixs}
\sim 
\mathrm{e}^{iax}\sum_{n=0}^\infty h^{(n)}(a)
\left( \frac{i}{x}\right)^{n+1} 
\quad\quad 
\mathrm{as}\,\,x\rightarrow\infty
\ .
\end{equation}
Taking $\cos(\omega s)=\mathrm{Re}\left( \mathrm{e}^{i \omega
s}\right)$, we apply this theorem to (\ref{inermasacc}) with $a=0$ and
$h(s)=1/s^2+1/{(\Delta \mathsf{x})}^2$, in which case $\epsilon
=2$. The expansion will thus proceed in inverse powers of $\omega^2$,
with coefficients given by $\tau$-derivatives
of~$\mathsf{x}(\tau)$. In the leading order we obtain
\begin{equation}
\label{largeomega}
\dot{F}_{\tau}(\omega)
=
-\frac{\omega}{2\pi}\Theta(-\omega)
+\frac{\ddot{\mathsf{x}} \cdot \dddot{\mathsf{x}}}{24 \pi^2 \omega^2}
+O\left( \omega^{-4}\right) 
\quad\quad 
\mathrm{as}\,\, \vert\omega\vert \rightarrow\infty
\ , 
\end{equation}
which shows that for a generic trajectory the first correction to the
inertial response is of order~$\omega^{-2}$. There exist however
trajectories for which the first correction is of higher order, owing
to the vanishing of some of the coefficients in~(\ref{theorem}). An
extreme example is the uniformly accelerated trajectory, for which
$\mathsf{x}^{(m)} \cdot \mathsf{x}^{(n)} = 0$ whenever $m+n$ is odd
and the coefficients of all inverse powers of $\omega^2$ vanish: The
asymptotic behaviour is in this case exponential, as seen
from~(\ref{planck}).

\subsection{Stationary trajectories} 
\label{subsec:stationary}

We shall now examine the consequences of our 
transition rate formula 
(\ref{resultado1-infty}) 
for the six families of \emph{stationary\/} motions, 
classified
in \cite{letaw}
and reviewed in~\cite{rosubookchapter,rosu}. 
These motions have the property that
${(\Delta\mathsf{x})}^2$ depends only on the proper time difference
between the two points along the trajectory.  As a consequence, the
transition rate of a detector switched on in the infinite past is
independent of the proper time.  We recall that as these motions are
precisely the orbits of timelike Killing vectors in Minkowski space,
they are the only motions in which an independent definition of
``particles" is available via the positive and negative frequency
decomposition with respect of a timelike Killing
vector~\cite{DavDrayMan,Srira-Pad,korsbak}.

We consider each of the six families in turn, 
characterising the family by the associated Killing vector. 
We shall in particular
verify that in each case both 
(\ref{eq:infty-bounds-lor}) 
and 
(\ref{eq:infty-bounds}) remain
bounded in the asymptotic past, justifying the limiting procedure that
led to (\ref{resultado1-infty}) for both of our detector models.

\subsubsection{Timelike translation} 

The first stationary trajectory is the inertial motion, 
$\mathsf{x}(\tau)=\mathsf{u}\tau$, 
where $\mathsf{u}$ is a constant timelike unit vector. 
The associated Killing vector generates a timelike translation. 
The non-inertial correction term in 
(\ref{inermasacc})
then vanishes and we obtain the expected result. 
It is evident that both (\ref{eq:infty-bounds-lor}) 
and 
(\ref{eq:infty-bounds}) remain bounded at $s\to\infty$. 

\subsubsection{Boost} 

The second stationary trajectory is the uniformly accelerated motion, 
or Rindler motion. 
The associated Killing vector generates a boost. 
Denoting the proper acceleration by~$a>0$, the trajectory reads 
\begin{equation}
\mathsf{x}(\tau)
=
\bigl( a^{-1} \sinh(a\tau), a^{-1} \cosh(a\tau), 0, 0\bigr) 
\ , 
\end{equation}
and we have 
${(\Delta \mathsf{x})}^2 = -4 a^{-2} \sinh^2(as/2)$. 
The tetrad vectors are 
\begin{equation}
\mathsf{e}_1=\bigl(\sinh(a\tau), \cosh(a\tau), 0, 0\bigr)
\ ,
\quad 
\mathsf{e}_2=(0,0,1,0)
\ ,
\quad
\mathsf{e}_3=(0,0,0,1)
\ . 
\end{equation}
The quantities 
(\ref{eq:infty-bounds-lor}) 
and 
(\ref{eq:infty-bounds}) remain bounded at $s\to\infty$ 
since the numerator 
and denominator in each diverges proportionally to~$\mathrm{e}^{as}$.

The spectrum was computed in \cite{schlicht,Schlicht:thesis} 
by performing the intergral over $s$ with $\epsilon>0$ 
and then taking the limit $\epsilon\to0$. 
To compute the spectrum directly from~(\ref{resultado1-infty}), 
we use the symmetry of the
integrand under $s\to-s$ to write
\begin{equation}
\label{eq:aux4}
\int_0^{\infty}\mathrm{d}s 
\left( -\frac{a^2\cos(\omega s)}{4\sinh^2(as/2)}
+\frac{1}{s^2}\right)
=
\frac{1}{2}
\int_{-\infty}^{\infty}\mathrm{d}s 
\left( -\frac{a^2 \cos(\omega s)}{4\sinh^2(as/2)}
+\frac{1}{s^2}\right)
\ . 
\end{equation}
We then deform the contour to the horizontal line
$\mathrm{Im}(s)=-\pi/a$. The second term in (\ref{eq:aux4}) gives a
vanishing contribution by contour integration, while the contribution
from the first term becomes \cite{gradshteyn}
\begin{equation}
\frac{a^2}{8}\cosh(\pi\omega/a)\int_{-\infty}^{\infty}\mathrm{d}s
\frac{\cos(\omega s)}{\cosh^2(as/2)}
=
\frac{\pi\omega}{2}\coth(\pi\omega/a)
\ . 
\end{equation}
Substituting this result in (\ref{resultado1-infty}) yields the
Planckian transition rate~(\ref{planck}). We have thus recovered the
Unruh effect.

\subsubsection{Timelike translation with rotation} 

The third stationary trajectory is the circular motion, given by 
\begin{equation}
\mathsf{x}(\tau)=
\bigl(\gamma \tau, R\cos(\gamma\Omega\tau), 
R\sin(\gamma\Omega\tau),0 \bigr)
\ , 
\end{equation}
where $R>0$, $\Omega\ne0$, $\vert R\Omega \vert <1$ and
$\gamma=(1-R^2\Omega^2)^{-1/2}$. The associated Killing vector is a
linear combination of a timelike translation generator and a rotation
generator. Solving the Fermi-Walker transport equations yields the
tetrad
\begin{align}
\mathsf{e}_1(\tau) 
& = 
\big(-\gamma\Omega R\sin(\gamma^2\Omega\tau),\, 
\cos(\gamma\Omega\tau)\cos(\gamma^2\Omega\tau)
+\gamma\sin(\gamma \Omega\tau)\sin(\gamma^2\Omega\tau),
\notag
\\ 
& 
\qquad\ 
\sin(\gamma\Omega\tau)\cos(\gamma^2\Omega\tau)
-\gamma\cos(\gamma \Omega\tau)\sin(\gamma^2\Omega\tau),\, 0 \big) 
\ , 
\notag
\\ 
\mathsf{e}_2(\tau) 
& = 
\big(\gamma\Omega R\cos(\gamma^2\Omega\tau),\, 
\cos(\gamma\Omega\tau)\sin(\gamma^2\Omega\tau)
-\gamma\sin(\gamma \Omega\tau)\cos(\gamma^2\Omega\tau), 
\notag
\\ 
& 
\qquad\ 
\sin(\gamma\Omega\tau)\sin(\gamma^2\Omega\tau)
+\gamma\cos(\gamma \Omega\tau)\cos(\gamma^2\Omega\tau),\, 0 \big) 
\ , 
\notag
\\ 
\mathsf{e}_3(\tau)
& = \big(0,0,0,1\big)
\ . 
\end{align}
At $s\to\infty$, $\Delta \mathsf{x}$ and ${(\Delta \mathsf{x})}^2$ grow 
respectively linearly and quadratically in~$s$, 
while $\dot{\mathsf{x}}$ and 
each $\mathsf{e}_i$ are bounded. 
Quantities (\ref{eq:infty-bounds-lor}) and 
(\ref{eq:infty-bounds}) therefore remain bounded at $s\to\infty$. 

The transition rate (\ref{resultado1-infty}) appears not known in
terms of elementary functions. Numerical results are given in  
\cite{letaw-pfautsch,letaw} and an analytic approximation 
in~\cite{mueller}. 

\subsubsection{Timelike translation with null rotation} 

The fourth stationary trajectory is the orbit of a Killing vector that
is a linear combination of a timelike translation generator and a null
rotation generator. The spatial projection is the planar curve
$y=\kappa \frac{\sqrt{2}}{3}x^{3/2}$, and the full trajectory reads
\begin{equation}
\mathsf{x}(\tau)
=
\bigl( 
\tau+\tfrac{1}{6}\kappa^2\tau^3,\,
\tfrac{1}{2}\kappa\tau^2,\,\tfrac{1}{6}\kappa^2\tau^3,\,0
\bigr) 
\ , 
\end{equation}
where $\kappa>0$. The tetrad is given by 
\begin{align}
\mathsf{e}_1(\tau) 
&= 
\big( \kappa\tau\cos(\kappa\tau)
+\tfrac12 \kappa^2\tau^2 \sin(\kappa\tau),\,
\cos(\kappa\tau)+\kappa\tau\sin(\kappa\tau),
\notag
\\
& 
\qquad
\kappa\tau\cos(\kappa\tau)
+(-1+ \tfrac12 \kappa^2\tau^2)\sin(\kappa\tau),\,0
\big) 
\ , 
\notag
\\
\mathsf{e}_2(\tau) 
&= 
\big( -\tfrac12 \kappa^2\tau^2\cos(\kappa\tau)
+\kappa\tau\sin(\kappa\tau),\,\kappa\tau\cos(\kappa\tau)
-\sin(\kappa\tau),
\notag
\\
& 
\qquad
(1-\tfrac12 \kappa^2\tau^2)\cos(\kappa\tau)+\kappa\tau\sin(\kappa\tau),\,0
\big) 
\ , 
\notag
\\
\mathsf{e}_3(\tau)
&=
(0,\,0,\,0,\,1)
\ , 
\end{align}
and it can be verified that quantities (\ref{eq:infty-bounds-lor})
and (\ref{eq:infty-bounds}) remain bounded at $s\to\infty$. The
transition rate can be computed from (\ref{inermasacc}) 
by the method of residues and equals 
\begin{equation}
\dot{F}_\tau(\omega)
=
- \frac{\omega}{2\pi}\Theta(-\omega)
+\frac{\kappa}{8\sqrt{3}\pi}\,
\mathrm{e}^{-2\sqrt{3}\vert\omega\vert/\kappa}
\ , 
\end{equation}
which agrees with the result given in~\cite{letaw}.

\subsubsection{Boost with spacelike translation} 

The fifth stationary trajectory is the orbit of a Killing vector that
is a linear combination of a boost
generator and a spacelike translation generator. 
The spatial projection is the catenary $y=k\cosh(y/b)$, and
the full trajectory reads
\begin{equation}
\mathsf{x}(\tau)=\frac{1}{a^2}\left( k\sinh(a\tau),\,
k\cosh(a\tau),\,ba\tau,0\right) 
\ , 
\end{equation}
where $a>0$, $b>0$ and $k = \sqrt{a^2+b^2}$. 
The tetrad vectors are 
\begin{align}
\mathsf{e}_1(\tau) 
& = 
\big( \sinh(a\tau)\cos(b\tau)+(b/a) \cosh(a\tau)\sin(b\tau), 
\notag
\\ 
&
\qquad 
\cosh(a\tau)\cos(b\tau)
+(b/a)\sinh(a\tau)\sin(b\tau),\,(k/a)\sin(b\tau),\, 0 \big) 
\ , 
\notag
\\ 
\mathsf{e}_2(\tau) 
& = 
\big( -\sinh(a\tau)\sin(b\tau)+(b/a)\cosh(a\tau)\cos(b\tau), 
\notag
\\ 
&
\qquad 
-\cosh(a\tau)\sin(b\tau)
+(b/a)\sinh(a\tau)\cos(b\tau),\,(k/a)\cos(b\tau),\, 0 \big) 
\ , 
\notag
\\ 
\mathsf{e}_3(\tau) 
& = 
(0,\,0,\,0,\,1)
\ . 
\end{align}
As in the uniformly accelerated case, quantities
(\ref{eq:infty-bounds-lor}) and (\ref{eq:infty-bounds}) 
remain bounded at $s\to\infty$ since the
numerator and denominator in each diverges proportionally
to~$\mathrm{e}^{as}$.  The transition rate (\ref{resultado1-infty})
appears not known in terms of elementary functions but numerical 
results are given in~\cite{letaw}. 

\subsubsection{Boost with rotation} 

The sixth stationary trajectory is the orbit of a Killing
vector that is a linear combination of a boost generator and a
rotation generator. The trajectory reads
\begin{equation}
\mathsf{x}(\tau)=\left( \frac{\alpha}{R_+}\sinh(R_+\tau),\,
\frac{\alpha}{R_+}\cosh(R_+\tau),\,\frac{\beta}{R_-}\cos(R_-\tau),\,
\frac{\beta}{R_-}\sin(R_-\tau)\right) 
\ , 
\end{equation}
where $R_\pm>0$, $\beta>0$ and $\alpha = \sqrt{1 + \beta^2}$. 
Defining $\Omega :=\sqrt{\alpha^2R_-^2+\beta^2R_+^2}$, 
the tetrad vectors can be written as 
\begin{align}
\mathsf{e}_1(\tau) 
& = 
\Omega^{-2} 
\Big(
\sinh(R_+\tau)\left[ \alpha^2 R_-^2+\beta^2 R_+\Omega\,
\sin(\Omega \tau)+\beta^2R_+^2\,\cos(\Omega\tau)\right], 
\notag
\\ 
& 
\qquad\qquad\!
\cosh(R_+\tau)\left[ \alpha^2 R_-^2+\beta^2 R_+\Omega\,
\sin(\Omega \tau)+\beta^2R_+^2\,\cos(\Omega\tau)\right],
\notag
\\ 
& 
\qquad\qquad\!
\alpha\beta R_+ \left[ R_- \bigl(1-\cos(\Omega\tau) \bigr) 
\cos(R_-\tau)-\Omega\sin(\Omega\tau)\sin(R_-\tau)\right]  ,
\notag
\\ 
& 
\qquad\qquad\!
\alpha\beta R_+ \left[ R_- \bigl(1-\cos(\Omega\tau) \bigr) 
\sin(R_-\tau)+\Omega\sin(\Omega\tau)\cos(R_-\tau)\right] \Big) 
\ , 
\notag
\\ 
\mathsf{e}_2(\tau) 
& =
\Omega^{-2}
\Big(
\alpha \beta R_- \left[ R_+ \bigl(1-\cos(\Omega\tau)\bigr) 
\sinh(R_+\tau)-\Omega\sin(\Omega\tau)\cosh(R_+\tau)\right]  ,
\notag
\\ 
& 
\qquad\qquad\!
\alpha\beta R_- \left[ R_+ \bigl(1-\cos(\Omega\tau)\bigr) 
\cosh(R_+\tau)-\Omega\sin(\Omega\tau)\sinh(R_-\tau)\right] ,
\notag
\\ 
& 
\qquad\qquad\!
\cos(R_-\tau) \bigl(\beta^2R_+^2+\alpha^2R_-^2\cos(\Omega\tau)\bigr)
+\alpha^2R_-\Omega\sin(R_-\tau)\sin(\Omega\tau),
\notag
\\ 
& 
\qquad\qquad\!
\sin(R_-\tau) \bigl(\beta^2R_+^2+\alpha^2R_-^2\cos(\Omega\tau)\bigr)
-\alpha^2R_-\Omega\cos(R_-\tau)\sin(\Omega\tau) 
\Big) 
\ , 
\notag
\\ 
\mathsf{e}_3(\tau) 
& = 
\Big(
\beta\left[ \cos(\Omega\tau)\cosh(R_+\tau)
-(R_+/\Omega)\sin(\Omega\tau)\sinh(R_+\tau)\right],
\notag
\\ 
& 
\qquad
\beta\left[ \cos(\Omega\tau)\sinh(R_+\tau)
-(R_+/\Omega)\sin(\Omega\tau)\cosh(R_+\tau)\right],
\notag
\\ 
& 
\qquad
\alpha \left[ -\cos(\Omega\tau)\sin(R_-\tau)  
-(R_-/\Omega)\sin(\Omega\tau)\cos(R_+\tau)\right],
\notag
\\ 
& 
\qquad
\alpha \left[ -\cos(\Omega\tau)\cos(R_-\tau)  
+(R_-/\Omega)\sin(\Omega\tau)\sin(R_+\tau)\right]
\Big) 
\ . 
\end{align}
Once again quantities (\ref{eq:infty-bounds-lor})
and (\ref{eq:infty-bounds}) remain bounded at
$s\to\infty$ since both the numerators and the denominators grow as
$\mathrm{e}^{R_+s}$. The transition rate (\ref{resultado1-infty})
appears not known in terms of elementary functions.

\subsection{From asymptotically inertial motion 
to asymptotically uniform acceleration}
\label{subsec:as-unruh} 

As a first example of nonstationary motion, we consider the trajectory 
\begin{equation}
\label{eq:interpolating-traj}
\mathsf{x}(\tau) = 
\Bigl(
\tau + {(2a)}^{-1}  
\bigl[
\mathrm{e}^{a\tau} - \ln(\mathrm{e}^{a\tau} + 1) 
\bigr]
\, , 
{(2a)}^{-1}  
\bigl[
\mathrm{e}^{a\tau} + \ln(\mathrm{e}^{a\tau} + 1) 
\bigr]
\, , 
0\, , 
0\, 
\Bigr)
\ , 
\end{equation}
where $a>0$. The magnitude of the proper acceleration is $a/ (1 +
\mathrm{e}^{-a\tau} )$. The trajectory is asymptotically inertial as
$\tau \to -\infty$ and has asymptotically uniform acceleration $a$ as
$\tau \to \infty$. As the quantities (\ref{eq:infty-bounds-lor}) 
and (\ref{eq:infty-bounds}) are
clearly bounded as $s\to\infty$, we are justified to use the
transition rate formulas (\ref{resultado1-infty}) and
(\ref{inermasacc}) for a detector switched on in the asymptotic past.

At large negative $\tau$, one expects the transition rate to asymptote
to that of inertial motion. To examine the noninertial term in
(\ref{inermasacc}) in this limit, we rearrange the integrand as 
\begin{align}
\label{eq:interpolating-aux1}
&
\frac{1}{{(\Delta \mathsf{x})}^2} 
+\frac{1}{s^2}
= 
\frac{- {(\Delta \mathsf{x})}^2 - s^2}{s^4}
{\left( 1 + \frac{- {(\Delta \mathsf{x})}^2 - s^2}{s^2}
\right)}^{-1}
\end{align} 
and note that 
(\ref{eq:interpolating-traj}) yields 
\begin{align}
\label{eq:interpolating-Delta2}
&
- {(\Delta \mathsf{x})}^2
= 
\frac{1}{a^2} 
\left[ 
as + \frac{g ( 1 - \mathrm{e}^{-as})}{(1-g)} 
\right] 
\left\{
as + \ln \bigl[ 1 - g ( 1 - \mathrm{e}^{-as}) \bigr] 
\right\}
\ , 
\end{align} 
where $g := 1/ (1 + \mathrm{e}^{-a\tau} )$. Expanding in
(\ref{eq:interpolating-Delta2}) the logarithm as $\ln(1-x) = -x -
\frac12 x^2 + O(x^3)$ shows that the second factor in
(\ref{eq:interpolating-aux1}) is of the form $1 + O(g^2)$, uniformly
in~$s$, and yields then in the first factor an estimate that can be
applied under the integral over~$s$. We find
\begin{align}
{\dot F}_{\tau}(\omega)
& =
-\frac{\omega}{2\pi}\Theta(-\omega)
+ \frac{a}{2\pi^2} h(\omega/a) 
\mathrm{e}^{2a\tau}
+ 
O(\mathrm{e}^{3a\tau})
\ , 
\ \ \ 
\tau\to-\infty
\ , 
\end{align}
where 
\begin{equation}
h(x) := 
\frac{1}{2} 
\int_0^\infty 
\mathrm{d}y
\, \frac{
(1-\mathrm{e}^{-y}) \bigl[ (2+y) \mathrm{e}^{-y} - 2 + y \bigr]
\cos(yx) 
}
{y^4}
\ . 
\end{equation}
An expression for $h$ in terms of elementary functions 
can be found by repeated integration by parts and use of
formulas 3.434 in~\cite{gradshteyn}. 
$h(x)$ is even in $x$ and
decreasing in~$x^2$, and it has the
asymptotic expansions $h(x) =
\frac13(1-\ln2)+ O\bigl(x^2 \ln(\vert x\vert)\bigr)$ 
as $x\to0$ and 
$h(x) = 1/(12x^2) + O \bigl(x^{-4}\bigr)$
as $\vert x\vert\to\infty$. 
The transition rate thus asymptotes to that of 
inertial motion as $\mathrm{e}^{2a\tau}$ when $\tau \to -\infty$. 

At large positive~$\tau$, one expects the transition rate to asymptote
to that of uniformly accelerated motion. We have verified from
(\ref{inermasacc}) that this is the case, using a monotone convergence
argument to take the limit under the integral, but we have not pursued
an estimate for the error term.

\subsection{From vanishing to diverging acceleration}
\label{subsec:div-acc} 

As the last example, we consider the two trajectories given by  
\begin{equation}
\label{eq:van-to-div}
\mathsf{x}(\tau)
=
\left( \frac{\tau^3}{6}-\frac{1}{2\tau},\,
\frac{\tau^3}{6}+\frac{1}{2\tau},\,0,\,0\right) 
\ , 
\end{equation}
one with $\tau\in(0,\infty)$ and the other with
$\tau\in(-\infty,0)$. 
The acceleration has magnitude $2/\vert\tau\vert$. 
The latter case was discussed
in~\cite{Schlicht:thesis}. 

Each of these trajectories resides entirely
in one Rindler quadrant. 
The trajectory with $\tau<0$ approaches 
$\scri^-$ as $\tau\to-\infty$ and 
$\scri^+$ as $\tau\to0_-$, 
having thus taken an infinite amount of proper time to come from 
$\scri^-$ in the past but reaching 
$\scri^+$ in finite proper time in the future. 
Note that this trajectory is not asymptotically inertial in the past
even though the proper acceleration tends to zero. 
The trajectory with 
$\tau>0$ is the time-reversed version, 
approaching 
$\scri^-$ as $\tau\to0_+$ and 
$\scri^+$ as $\tau\to\infty$ 
and having taken a finite amount of proper time to come from 
$\scri^-$ in the past. 

The tetrad vectors are 
\begin{equation}
\mathsf{e}_1(\tau)=\left(\frac{\tau^2}{2}-\frac{1}{2\tau^2},\, 
\frac{\tau^2}{2}+\frac{1}{2\tau^2},\, 0,\, 0\right),
\quad 
\mathsf{e}_2(\tau)=(0,0,1,0),
\quad
\mathsf{e}_3(\tau) =(0,0,0,1)
\ . 
\end{equation}
On the trajectory with $\tau>0$, 
both of quantities (\ref{eq:infty-bounds-lor}) 
and the first and third of the quantities
(\ref{eq:infty-bounds}) diverge as $s\to\tau$. 
The limiting arguments of subsections \ref{subsec:lor-as}
and 
\ref{subsec:taunought-infty}
do therefore not justify our formula for a detector turned on in the
asymptotic past (of finite proper time) for this trajectory. 
We have not investigated whether an improved set of limiting arguments
could be found. 

On the trajectory with $\tau<0$, the quantities 
(\ref{eq:infty-bounds-lor}) 
and (\ref{eq:infty-bounds}) all 
remain bounded as $s\to\infty$, and we are
justified to use formula 
(\ref{inermasacc}) for a detector switched on in the asymptotic
past. We find 
\begin{align}
{\dot F}_{\tau}(\omega)
& =
-\frac{\omega}{2\pi}\Theta(-\omega)
+\frac{1}{2\pi^2(-\tau)}
\int_0^{\infty}\textrm{d}y
\, 
\frac{\cos (\omega \tau y)}{y^2 + 3y + 3}
\ , 
\label{eq:divaccfut-result}
\end{align}
where we have introduced the new integration variable $y:=s/(-\tau)$. 
For $\omega=0$, the noninertial term in 
(\ref{eq:divaccfut-result}) equals 
${\bigl[6\pi \sqrt{3} (-\tau)\bigr]}^{-1}$. 
For $\omega\ne0$, the 
asymptotic behaviour at $\tau\to-\infty$ can be found using 
(\ref{theorem}) and that at $\tau\to0_-$ by techniques similar to
those in section~\ref{sec:lor-limit}. 
The result is 
\begin{subequations}
\label{eq:divaccfut-as}
\begin{align}
{\dot F}_{\tau}(\omega)
& =
-\frac{\omega}{2\pi}\Theta(-\omega)
+ \frac{3}{2\pi^2 \omega^2 {(-\tau)}^3} 
\left[ 
1 + O\left(\frac{1}{\omega^2\tau^2}\right)
\right] 
\ , 
\ \ \ 
\tau\to-\infty
\ , 
\\
\noalign{\medskip}
{\dot F}_{\tau}(\omega)
& =
\frac{1}{6\pi \sqrt{3} (-\tau)} 
-\frac{\omega}{4\pi}
- \frac{3 \vert\omega\vert }{4\pi^2}
\Bigl[\vert\omega\tau\vert
\ln(\vert\omega\tau\vert)
+ O\bigl(\vert\omega\tau\vert\bigr)
\Bigr] 
\ , 
\ \ \ 
\tau\to0_-
\ . 
\end{align}
\end{subequations}
The noninertial contribution to the transition rate thus vanishes as
${(-\tau)}^{-3}$ when $\tau\to-\infty$ for $\omega\ne0$ but diverges
as ${(-\tau)}^{-1}$ when $\tau\to0_-$.  This is consistent with what
one might have expected from the magnitude of the proper acceleration
in these limits.

\section{Conclusions}
\label{sec:conclusions}

We have analysed a particle detector model whose coupling to a
massless scalar field in four-dimensional Minkowski space is
regularised by a spatial profile, rigid in the detector's
instantaneous rest frame. When the profile is given by the Lorentzian
function as in (\ref{eq:profile-scaling}) and~(\ref{lorentzian}), we
computed explicitly the zero-size limit of the instantaneous
transition rate, obtaining a manifestly finite integral formula that
no longer involves regulators or limits. We then considered a detector
model with a modified definition of spatial smearing and showed, under
certain technical conditions, that the instantaneous transition rate
is independent of the choice of the profile function and agrees with
that obtained from the (unmodified) smearing with the Lorentzian
profile. 
The formulas for the 
transition rate in the cases of 
finite and infinite proper time of detection are, respectively, 
\begin{equation}
\label{resultado1-conc}
\dot{F}_{\tau}(\omega)
=
-\frac{\omega}{4\pi}+\frac{1}{2\pi^2}
\int_0^{\Delta\tau}\textrm{d}s
\left( 
\frac{\cos (\omega s)}{{(\Delta \mathsf{x})}^2} 
+ 
\frac{1}{s^2} 
\right) 
\ \ +\frac{1}{2\pi^2 \Delta \tau}
\ , 
\ \ \Delta \tau> 0
\ , 
\end{equation}
and 
\begin{align}
\dot{F}_{\tau}(\omega)
&=
-\frac{\omega}{4\pi}+\frac{1}{2\pi^2}
\int_0^{\infty}\textrm{d}s
\left( 
\frac{\cos (\omega s)}{{(\Delta \mathsf{x})}^2} 
+ 
\frac{1}{s^2} 
\right) 
\notag
\\
\noalign{\vspace\jot}
& = 
-\frac{\omega}{2\pi}\Theta(-\omega)
+\frac{1}{2\pi^2}\int_0^{\infty}\textrm{d}s\cos (\omega s)
\left( 
\frac{1}{{(\Delta \mathsf{x})}^2} 
+\frac{1}{s^2}
\right) 
\ . 
\label{resultado1-infty-conc}
\end{align}
For a detector switched on in the asymptotic past, we have justified
these formulas from our detector models under the assumption that
quantities (\ref{eq:infty-bounds-lor}) or (\ref{eq:infty-bounds})
remain bounded in the asymptotic past. We have not examined whether
this boundedness condition could be relaxed.

We showed that the acceleration only affects the part of the
transition rate that is even as a function of the frequency~$\omega$,
and we obtained a method to compute the coefficients of all inverse
powers of $\omega^2$ in the asymptotic large $\omega^2$ expansion of
the transition rate. Finally, we applied our transition rate formula
to a number of examples, including all stationary trajectories. We
recovered in particular the Unruh effect for uniform acceleration, and
we obtained an interpolating transition rate for a nonstationary
trajectory that interpolates between asymptotically inertial and
asymptotically uniformly accelerated motion. We believe that these
results strongly support the use of (\ref{resultado1-conc}) and
(\ref{resultado1-infty-conc}) as defining the instantaneous transition
rate of an Unruh particle detector. 

We re-emphasise that the need for a spatial smearing arose because we
chose to address the instantaneous transition rate of the detector
\emph{while the interaction is switched on\/}, rather than the total
excitation probability after the interaction has been smoothly
switched on and off. The formal first-order perturbation theory
expression for the transition rate involves in this case the field's
Wightman function in a way that is ill-defined without
regularisation. As first observed by Schlicht for the Rindler
trajectory~\cite{schlicht}, and as we have verified in appendix
\ref{app:ieps}
for arbitrary noninertial trajectories, the conventional $i\epsilon$
regularisation of the Wightman function in a given Lorentz frame
results into a Lorentz-noninvariant expression for the transition rate
and is hence not viable. The regularisation by a spatial profile, by
contrast, is manifestly Lorentz invariant.

Our modified detector model of section \ref{sec:generalprofile} was
related to spatial smearing with a profile function of compact
support, and we showed that the transition rate in this model is
independent of the profile function. We did not demonstrate the model
to be equivalent to spatial smearing with a profile function of
compact support, owing to the possibility that the integration over
the spatial surfaces as defined in (\ref{eq:smearedFdot-def}) could
miss a distributional part of the integrand. However, as the modified
detector model yields in the zero-size limit 
the same transition rate as spatial smearing
with the Lorentzian profile function (which is not of compact
support), we suspect the model to be equivalent to spatial
smearing for at least some classes of profile
functions. This question would deserve further study.

An alternative to spatial smearing could be to define the
instantaneous transition rate by starting with a pointlike detector
and smooth switching function of compact support and then taking a
limit in which the switching function approaches the characteristic
function of an interval. We are not aware of reasons to expect an
unambiguous limit to exist, but might there be specific limiting
prescriptions that reproduce the result obtained with spatial
smearing?

With a sharp switch-on at the initial time $\tau_0$, the transition
rate (\ref{resultado1-conc}) diverges as ${(\tau-\tau_0)}^{-1}$ when
$\tau\to\tau_0$. The total transition probability, obtained by
integrating the transition rate, is therefore infinite, owing to the
violent switch-on event, regardless how small the coupling constant in
the interaction Hamiltonian is. For the stationary trajectories the
transition rate (\ref{resultado1-infty-conc}) of a detector switched on in
the asymptotic past is constant in time, and the total transition
probability is again infinite, now owing to the infinite amount of
time elapsed in the past. In these situations one may therefore have
reason to view our results, all of which were obtained within
first-order perturbation theory, as suspect, and perhaps even to
question the whole notion of the instantaneous transition
rate. However, in situations where the detector is switched on in the
asymptotic past of infinite proper time and the total probability of
excitation ($\omega>0$) is finite, the first-order perturbation theory
result should be reliable at least for the excitation rate, although
the total probability of de-excitation ($\omega<0$) then still
diverges. Two examples of this situation, 
with an acceleration that vanishes asymptotically in the past, 
were found in
section~\ref{sec:applications}.

When the total excitation probability is finite, it has a directly
observable meaning as the fraction of detectors that have become
excited in an ensemble that is initially prepared in the state
$\vert0\rangle_d$ and follows the trajectory~$\mathsf{x}(\tau)$. Note,
however, that observing the ensemble changes the initial conditions
for the subsequent dynamics, and a single ensemble can thus be used to
measure the excitation probability at only one value of proper
time. To measure the excitation probability at several values of the
proper time requires a family of identically prepared ensembles, each
of which will be used to read off the excitation probablility at one
value of the proper time only. The excitation rate, proportional to
$\dot{F}_{\tau}(\omega)$, is then the proper time derivative of this
probability. Relating $\dot{F}_{\tau}(\omega)$ to the acceleration
effects that may become practically observable in particle
accelerators \cite{bell-leinaas} remains thus a subtle issue
\cite{rosubookchapter,rosu}.\footnote{We thank 
Hans Westman for discussions on this point.} 

It would be interesting to investigate to what extent our results can
be generalised to the variety of situations to which Schlicht's
Lorentzian profile detector was generalised in
\cite{Langlois,Langlois-thesis}. For example, do the formulas
(\ref{resultado1-conc}) and (\ref{resultado1-infty-conc}) 
generalise to
spacetime dimensions other than four, and if yes, what is the form of
the subtraction term? Does the clean separation of the spectrum into its
even and odd parts continue? 
Further, to what extent can the notion of spatial profile be employed 
to regularise the transition rate in a curved spacetime, 
presumably reproducing known results for stationary trajectories 
\cite{sonego-westman} but also allowing nonstationary motion? 
In particular, might there be a connection with the
regularisation prescriptions of the classical self-force
problem~\cite{DetWhit,poisson-livrev,poisson-gr17,anderson-wiseman}? 
Finally, would a nonperturbative treatment be
feasible?

\section*{Acknowledgements}
We thank 
John Barrett, 
Kirill Krasnov, 
Paul Langlois, 
Hans Westman, 
Bernard Whiting  
and especially 
Chris Fewster 
for helpful discussions, 
Haret Rosu and Douglas Singleton for 
bringing their related work to our attention 
and an anonymous referee, Alan Wiseman and Eric Poisson for 
locating references. 
JL acknowledges hospitality and financial support 
of the Isaac Newton Institute programme 
``Global Problems in Mathematical Relativity"
and of the Perimeter Institute for Theoretical Physics. 
AS was supported by an EPSRC Dorothy Hodgkin 
Research Award to the University of Nottingham.

\begin{appendix}

\section{Appendix: $i\epsilon$ regularisation in a given Lorentz frame}
\label{app:ieps}

In this appendix we evaluate the $\epsilon\to0$ limits of the
instantaneous transition rates (\ref{defexcitation-sharp}) and
(\ref{defexcitation-sharp-infty}) for for the $i\epsilon$ regularisation
(\ref{tradWightman}) in a given Lorentz frame. 
The results will differ from 
(\ref{resultado1}) and (\ref{resultado1-infty})
by an additive Lorentz-noninvariant term. 
This generalises observations of Schlicht     
in the special case of uniformly accelerated motion \cite{schlicht} 
and supports the view that the $i\epsilon$ regularisation 
is physically inappropriate in the context of 
instantaneous
transition rate calculations.

The notation follows subsection~\ref{subsec:lor-sharp}. 

We start with finite $\tau_0$ and assume the trajectory to be $C^9$ in
the closed interval $[\tau_0, \tau]$. With the $i\epsilon$ regularised
correlation function~(\ref{tradWightman}), the transition rate for a
detector switched on at $\tau_0$ reads
\begin{equation}
\label{Fieps}
\dot{F}_{\tau} (\omega)
=
\frac{1}{2\pi^2}\ 
\mathrm{Re}
\int_{0}^{\Delta\tau}\mathrm{d}s\,
\frac{\mathrm{e}^{-i\omega s}}
{
{\vert\mathbf{x}(\tau)-\mathbf{x}(\tau-s)\vert}^2
- {\bigl[t(\tau)-t(\tau-s)-i\epsilon\bigr]}^2
}
\ . 
\end{equation}
The decomposition of $\dot{F}_\tau(\omega)$ 
into its even and odd parts in $\omega$
is obtained from formulas 
(\ref{eq:even-and-odd}) 
with the replacements 
\begin{equation}
\label{eq:ieps-replace}
\vsigma^2 \to -1, 
\ \ 
\vsigma\cdot\Delta \mathsf{x} \to - \Delta t
\ , 
\end{equation}
where $\Delta t := t(\tau)-t(\tau-s)$. After the further decomposition
into contributions from the integration subintervals $s \in
\bigl[0,\sqrt{\epsilon}\bigr]$ and $s \in \bigl[\sqrt{\epsilon},
\Delta\tau\bigr]$, the estimates of subsection \ref{subsec:lor-sharp}
readily adapt to show that the contribution from the latter
subinterval is again given by~(\ref{eq:large-even-and-odd}).

In the subinterval $s \in \bigl[0,\sqrt{\epsilon}\bigr]$, 
the denominators 
have now the expansion 
\begin{equation}
\label{eq:den-ieps-finalexpansion}
{\bigl[\epsilon^2 + {(\Delta\mathsf{x})}^2\bigr]}^2
+4\epsilon^2 {(\Delta t)}^2
= 
\epsilon^2 P
\left[ 
1 + \tfrac16  {\ddot{\mathsf{x}}}^2
\epsilon r^2 
- 4 \dot{t} \ddot{t} P^{-1} \epsilon^{3/2} r^3 
+ O(\epsilon^{3/2}) 
\right] 
\ , 
\end{equation}
where $s = \sqrt{\epsilon} r$ and 
$P := \epsilon^2+2(2\dot{t}^2-1)\epsilon r^2+r^4$. Note that 
the term $P^{-1} \epsilon^{3/2} r^3$ is of order 
$O(\epsilon)$, uniformly in $r$. 

Keeping in $\dot{F}_{<}^{\mathrm{odd}}$ just the leading term in 
(\ref{eq:den-ieps-finalexpansion}) and in the numerator 
just the leading power of~$s$, we find 
\begin{align}
\dot{F}^{\mathrm{odd}}_<
&= 
- \frac{\omega \dot{t} \sqrt{\epsilon}}{\pi^2}
\int_{0}^{1}
\mathrm{d}r 
\, 
\frac{r^2}{P}
\bigl[1 + O(\epsilon) \bigr]
\notag
\\
&= 
- \frac{\omega}{4\pi}
+ O(\sqrt{\epsilon})
\ , 
\label{eq:small-ieps-odd}
\end{align}
where the integral is elementary. 
In $\dot{F}_{<}^{\mathrm{even}}$ 
we use (\ref{eq:den-ieps-finalexpansion})
in the denominator and expand the numerator 
to next-to-leading order in~$s$, with the result  
\begin{align}
\dot{F}^{\mathrm{even}}_<
&= 
\frac{1}{2 \pi^2 \sqrt{\epsilon}}
\int_{0}^{1}
\mathrm{d}r 
\, 
\frac{(\epsilon - r^2)
\left[
1 
+ 4 \dot{t} \ddot{t} P^{-1} \epsilon^{3/2} r^3 
+ r^4 O(\epsilon) 
+ r^3 O(\epsilon^{3/2}) 
+ O(\epsilon^{5/2}) 
\right]}
{P}
\notag
\\
&=
\frac{1}{2 \pi^2 \sqrt{\epsilon}}
-\frac{1}{4\pi^2}\frac{\ddot{t}}{{(\dot{t}^2-1)}^{3/2}}
\left[ \dot{t}\sqrt{\dot{t}^2-1}
+\ln\!\left(\dot{t}-\sqrt{\dot{t}^2-1}\,\right)\right] 
+ O(\sqrt{\epsilon}) 
\ , 
\label{eq:small-ieps-even}
\end{align}
where the integrals are elementary, 
and in the last expression the 
term involving $\ddot{t}$ should for 
$\dot{t}=1$ be understood as its limiting value~$0$. . 

Combining these results and taking the limit $\epsilon\to0$, 
we find that the
transition rate differs from (\ref{resultado1}) 
by the additive term 
\begin{equation}
\label{nonLorentz-additive}
-\frac{1}{4\pi^2}\frac{\ddot{t}}{{(\dot{t}^2-1)}^{3/2}}
\left[ \dot{t}\sqrt{\dot{t}^2-1}
+\ln\!\left(\dot{t}-\sqrt{\dot{t}^2-1}\,\right)\right] 
\ , 
\end{equation}
understood for $\dot{t}=1$ as its limiting value~$0$. The term
(\ref{nonLorentz-additive}) clearly vanishes for inertial
trajectories. Given a point at which the proper acceleration is
nonzero, (\ref{nonLorentz-additive}) vanishes in Lorentz frames in
which $\partial_t$ is the velocity but is nonvanishing in
Lorentz frames in which $\partial_t$ is in the plane spanned by the
velocity and the acceleration but not proportional to the
velocity. The term (\ref{nonLorentz-additive}) is therefore Lorentz
invariant only for inertial trajectories.

Finally, these observations generalise to the switch-on in the
asymptotic past provided the trajectory is asymptotically sufficiently
well-behaved. From (\ref{eq:infty-bounds-lor}) and
(\ref{eq:ieps-replace}) it is seen that a sufficient condition is that
$(\Delta t) / {(\Delta \mathsf{x})}^2$ remains bounded as $s$
increases. This condition is in particular satisfied for uniformly
accelerated motion. 
We have verified that the analytical and numerical results given in
\cite{schlicht} for uniformly accelerated motion are consistent
with the sum of the Planckian spectrum (\ref{planck}) and the
Lorentz-noninvariant term~(\ref{nonLorentz-additive}).

\section{Appendix: 
Integral estimates for section \ref{sec:generalprofile}}
\label{app:contourestimates}

In this appendix we provide the required 
estimates for the integral (\ref{eq:GfromH}) 
over the contour~(\ref{eq:contour}).
We write $\auxeps := \sqrt{\delta}$.

\subsection{$C_1$}

We parametrise $C_1$ as in~(\ref{eq:contour}), 
$s = -ir$ with $0\leq r \leq\auxeps$. 
Let 
\begin{align}
\label{eq:R-def}
R(s) 
& := 
s_0^2 + H_2 s^2
\ . 
\end{align}
As it follows from 
(\ref{eq:SA-small-coeffs}) that 
$R>0$ for sufficiently small~$\auxeps$, we may write 
the contribution to (\ref{eq:GfromH}) from $C_1$ as 
\begin{equation}
\label{eq:GC1-div1}
{G}_{\tau,\tau_0}^{C_1} 
(\boldsymbol{\xi} , \boldsymbol{\xi}' ; \omega)
= 
\frac{1}{2\pi^2}
\, \mathrm{Im}\int_0^{\auxeps} 
\, \mathrm{d}r \, 
\frac{\mathrm{e}^{-\omega r}}{R \bigl[1 + (\mathsf{H}^2-R)/R\bigr]}
\ . 
\end{equation}
As $r^2/R = O(\auxeps^0)$ and the small $s$ expansion of 
$\mathsf{H}^2-R$ starts with $s^3$, we may approximate the factor 
${\bigl[1 + (\mathsf{H}^2-R)/R\bigr]}^{-1}$ 
in (\ref{eq:GC1-div1})
by the first two terms in its geometric 
series expansion, at the expense of an 
error of order $O(\auxeps)$ in the integral in~(\ref{eq:GC1-div1}). 
The contribution from the zeroth-order term 
vanishes on taking the imaginary part. In the first order term 
we may replace $\mathsf{H}^2-R$ by its $s^3$ term and the factor 
$\mathrm{e}^{-\omega r}$ by $1$ at the expense of 
an error of order $O(\auxeps)$ in the integral. We thus obtain 
\begin{equation}
\label{eq:GC1-div2}
{G}_{\tau,\tau_0}^{C_1} 
(\boldsymbol{\xi} , \boldsymbol{\xi}' ; \omega)
= 
- \frac{H_3}{2\pi^2}
\, \int_0^{\auxeps} 
\, \mathrm{d}r \, 
\frac{r^3}{{\bigl[s_0^2 - H_2 r^2 \bigr]}^2}
\ \ \ 
+ O(\auxeps)
\ . 
\end{equation}
Performing the elementary integral and using $s_0 = O(\auxeps^2)$, we
find
\begin{equation}
\label{eq:GC1-final}
{G}_{\tau,\tau_0}^{C_1} 
(\boldsymbol{\xi} , \boldsymbol{\xi}' ; \omega)
= 
\frac{H_3}{2\pi^2 H_2^2}
\ln(s_0) 
\ \ + O(\auxeps)
\ . 
\end{equation}
As the coefficient of $\ln(s_0)$ 
in (\ref{eq:GC1-final}) is of order~$O(\auxeps^2)$, 
(\ref{eq:GSA-C1:2}) follows. 

We note that for the uniformly accelerated trajectory the coefficient
of $\ln(s_0)$ in (\ref{eq:GC1-final}) vanishes, since for this
trajectory $\dddot{\mathsf{x}}$ is proportional to $\dot{\mathsf{x}}$
and $H_3$ hence vanishes by~(\ref{eq:SA-small-coeffs}).

\subsection{$C_2 \cup C_3 \cup C_4$}

On $C_2 \cup C_3$, it follows from 
(\ref{eq:contour}) 
and 
$0 < s_0 < 2 \delta = 2 \auxeps^2$
that 
$\vert - s^2 + s_0^2 \vert > \tfrac12 \auxeps^2$ 
for sufficiently small~$\auxeps$. 
Using 
(\ref{eq:SA-small}) and~(\ref{eq:SA-small-coeffs}), 
we may thus write 
the integrand in 
(\ref{eq:GfromH}) as 
\begin{equation}
\label{eq:aux1}
\frac{\mathrm{e}^{-i\omega s}}{\mathsf{H}^2}
=
\frac{1 -i\omega s}{s_0^2 -s^2} 
+ O(\auxeps^0)
\ . 
\end{equation}
The integral of the first term in (\ref{eq:aux1})
is elementary and the integral of the 
second term is of order~$O(\auxeps)$. 
We obtain 
\begin{equation}
\label{eq:GSA-C2C3}
{G}_{\tau,\tau_0}^{C_2 \cup C_3} 
(\boldsymbol{\xi} , \boldsymbol{\xi}' ; \omega)
= 
\frac{1}{4\pi^2s_0}
\ln\left( \frac{\auxeps+s_0}{\auxeps-s_0}\right) -\frac{\omega}{4\pi}
+O(\auxeps)
\ . 
\end{equation}

Consider then $C_4$ under the assumptions of
subsection~\ref{subsec:G}: $\tau_0$ is finite and the trajectory is
defined in the closed proper time interval $[\tau_0, \tau]$. 
We add and subtract $1/(s^2 - s_0^2)$ 
in the integrand in~(\ref{eq:GfromH}), obtaining 
\begin{align}
{G}_{\tau,\tau_0}^{C_4} 
(\boldsymbol{\xi} , \boldsymbol{\xi}' ; \omega)
& = 
\frac{1}{2\pi^2}\int_\auxeps^{\Delta \tau} 
\mathrm{d}s
\, 
\left(
\frac{\cos(\omega s)}{\mathsf{H}^2} 
+  \frac{1}{s^2 - s_0^2}
\right)
\ , 
\notag
\\
\noalign{\smallskip}
& 
\ \ \ \ 
-\frac{1}{4\pi^2s_0}\ln\left( \frac{\auxeps+s_0}{\auxeps-s_0}\right) 
+\frac{1}{4\pi^2s_0}\ln\left( \frac{\Delta\tau+s_0}{\Delta\tau-s_0}\right) 
\ , 
\label{eq:GSA-C4:1}
\end{align}
where we have 
evaluated the integral of the subtraction term. 
The second logarithm term in (\ref{eq:GSA-C4:1}) 
is equal to ${(2\pi^2 \Delta \tau)}^{-1}$ plus a correction 
of order~$O(\auxeps^4)$. Adding 
(\ref{eq:GSA-C2C3}) and~(\ref{eq:GSA-C4:1}), the 
$\auxeps$-dependent logarithm terms cancel and we obtain 
\begin{equation}
{G}_{\tau,\tau_0}^{C_2 \cup C_3 \cup C_4} 
(\boldsymbol{\xi} , \boldsymbol{\xi}' ; \omega)
= 
-\frac{\omega}{4\pi}
+ 
\frac{1}{2\pi^2}\int_\auxeps^{\Delta \tau} 
\mathrm{d}s
\, 
\left(
\frac{\cos(\omega s)}{\mathsf{H}^2} 
+  \frac{1}{s^2 - s_0^2}
\right)
\ \ +\frac{1}{2\pi^2 \Delta \tau}
+ O(\auxeps)
\ . 
\label{eq:GSA-C2C3C4}
\end{equation}

Let $I$ denote $2\pi^2$ times the integral term 
in~(\ref{eq:GSA-C2C3C4}). 
Adding and subtracting in the integrand its limiting 
value as $(\boldsymbol{\xi}, \boldsymbol{\xi}') \to 
(\boldsymbol{0}, \boldsymbol{0})$, 
we obtain the rearrangement 
$I = I_1 + I_2 + I_3$, where 
\begin{subequations}
\label{eq:Is} 
\begin{align}
I_1
& := 
\int_\auxeps^{\Delta \tau} 
\mathrm{d}s
\, 
\left(
\frac{\cos (\omega s)}{{(\Delta \mathsf{x})}^2} 
+  \frac{1}{s^2}
\right)
\ ,
\label{eq:I1}
\\
\noalign{\smallskip}
I_2
& := 
\int_\auxeps^{\Delta \tau} 
\mathrm{d}s
\, 
\left(
\frac{1}{(s^2 - s_0^2)}
-\frac{1}{s^2}
\right)
\ , 
\label{eq:I2}
\\
\noalign{\smallskip}
I_3
& := 
\int_\auxeps^{\Delta \tau} 
\mathrm{d}s
\, 
\cos(\omega s) 
\left(
\frac{1}{\mathsf{H}^2} 
- 
\frac{1}{{(\Delta \mathsf{x})}^2} 
\right)
\ .
\label{eq:I3}
\end{align}
\end{subequations}
In~$I_1$, 
it follows from (\ref{eq:small-s-Deltax}) that the integrand has a
small $s$ expansion that starts with a constant term, and the lower
limit can hence be replaced by zero at the expense of an error of
order~$O(\auxeps)$. $I_2$~is elementary and of
order~$O(\auxeps)$. In~$I_3$,
we rearrange the integrand as 
\begin{equation}
\label{eq:integrand3-rearr}
- \frac{\cos(\omega s)}{{(\Delta \mathsf{x})}^2}
\left[
1 - 
\frac{1}{1 + 
{\displaystyle 
\frac{\mathsf{H}^2 - {(\Delta \mathsf{x})}^2}{{(\Delta \mathsf{x})}^2}}}
\right]
\ . 
\end{equation}
It follows from~(\ref{eq:small-s-Deltax}), 
(\ref{eq:SA-small}), (\ref{eq:SA-small-coeffs}) 
and the inequalities 
$s_0 < 2\auxeps^2$ 
and 
$s^2 \leq \left\vert {(\Delta \mathsf{x})}^2
\right\vert$ 
that the combination 
$\left[ \mathsf{H}^2 - {(\Delta \mathsf{x})}^2 \right] 
/ \bigl[{(\Delta \mathsf{x})}^2\bigr]$ is of order $O(\auxeps^2)$. Hence 
(\ref{eq:integrand3-rearr}) is of the form~$O(\auxeps^2)/s^2$, 
from which we obtain $I_3 = O(\auxeps)$. 
Substituting these observations
in~(\ref{eq:GSA-C2C3C4}), equation
(\ref{eq:GSA-C2C3C4:2}) follows. 

Consider finally $C_4$ under the assumptions of
subsection~\ref{subsec:taunought-infty}: Either the trajectory is
defined for arbitrarily negative proper times and $\Delta\tau$ is
replaced by infinity, or the trajectory has come from infinity in
finite proper time and $\tau_0$ is understood as the (asymptotic)
value of the proper time at which the trajectory starts out at
infinity. In either case the only nontrivial change occurs in that now
there is a need to control the integrand in~(\ref{eq:I3}), given
by~(\ref{eq:integrand3-rearr}), also as $s$ increases respectively to
$\infty$ or to $\tau -\tau_0$. A~sufficient condition for this control
is to assume that the quantities (\ref{eq:infty-bounds}) all remain
bounded as $s$ increases. Under this assumption it follows from
(\ref{eq:H2-split})--(\ref{eq:SA-small-coeffs}) that
$\mathsf{H}^2 / \left[{(\Delta \mathsf{x})}^2\right] 
= 1+O(\auxeps^2)$, uniformly 
in~$s$. Hence 
(\ref{eq:integrand3-rearr}) is again of the form 
$O(\auxeps^2)/s^2$, and we obtain 
$I_3 = O(\auxeps)$.

\end{appendix}

\end{document}